\newif\iffigs\figstrue
\documentclass[12pt]{article}
\usepackage{latexsym,amssymb,lscape}
\usepackage{graphicx}        
\newtheorem{congettura}{Conjecture}[section]
\newtheorem{definizione}{Definition}[section]

\newtheorem{statement}{Statement}[section]

\def\IC{\relax\,\hbox{$\inbar\kern-.3em{\rm C}$}}
\def\IG{\relax\,\hbox{$\inbar\kern-.3em{\rm G}$}}
\def\IB{\relax{\rm I\kern-.18em B}}
\def\ID{\relax{\rm I\kern-.18em D}}
\def\IL{\relax{\rm I\kern-.18em L}}
\def\IF{\relax{\rm I\kern-.18em F}}
\def\IH{\relax{\rm I\kern-.18em H}}
\def\II{\relax{\rm I\kern-.17em I}}
\def\IN{\relax{\rm I\kern-.18em N}}
\def\IP{\relax{\rm I\kern-.18em P}}
\def\IQ{\relax\,\hbox{$\inbar\kern-.3em{\rm Q}$}}
\def\bfzero{\relax\,\hbox{$\inbar\kern-.3em{\rm 0}$}}
\def\IK{\relax{\rm I\kern-.18em K}}
\def\IG{\relax\,\hbox{$\inbar\kern-.3em{\rm G}$}}
 \font\cmss=cmss10 \font\cmsss=cmss10 at 7pt
\def\IR{\relax{\rm I\kern-.18em R}}
\def\ZZ{\relax\ifmmode\mathchoice
{\hbox{\cmss Z\kern-.4em Z}}{\hbox{\cmss Z\kern-.4em Z}}
{\lower.9pt\hbox{\cmsss Z\kern-.4em Z}} {\lower1.2pt\hbox{\cmsss
Z\kern-.4em Z}}\else{\cmss Z\kern-.4em Z}\fi}
\def\bfone{\relax{\rm 1\kern-.35em 1}}

\def\diag{{\rm diag}}

\def\Solv{\mathop{\rm Solv}\nolimits}
\def\Riem{\mathop{\rm Riem}\nolimits}
\def\inbar{\vrule height1.5ex width.4pt depth0pt}
\def\bfzero{\relax{\rm I\kern-.18em 0}}
\def\bfone{\relax{\rm 1\kern-.35em 1}}

\newcommand{\ft}[2]{{\textstyle\frac{#1}{#2}}}

\def\1bar{1\hskip -.275cm -}
\def\2bar{2\hskip -.275cm -}
\def\3bar{3\hskip -.275cm -}

\newsavebox{\uuunit}
\sbox{\uuunit}
                 {\setlength{\unitlength}{0.825em}
                      \begin{picture}(0.6,0.7)
                                      \thinlines
                                      \put(0,0){\line(1,0){0.5}}
                                      \put(0.15,0){\line(0,1){0.7}}
                                      \put(0.35,0){\line(0,1){0.8}}
                                     \multiput(0.3,0.8)(-0.04,-0.02){10}{\rule{0.5pt}{0.5pt}}
                      \end {picture}}

\makeatletter \@addtoreset{equation}{section} \makeatother

\newcommand{\be}{\begin{equation}}
\newcommand{\ee}{\end{equation}}
\newcommand{\ba}{\begin{eqnarray}}
\newcommand{\ea}{\end{eqnarray}}
\def\bfone{\relax{\rm 1\kern-.35em 1}}

\def\bfone{\relax{\rm 1\kern-.35em 1}}
\font\cmss=cmss10 \font\cmsss=cmss10 at 7pt

\newcommand{\so}{\mathfrak{so}}
\newcommand{\su}{\mathfrak{su}}
\newcommand{\usp}{\mathfrak{usp}}

\newcommand{\sym}{\mathfrak{sp}}
\newcommand{\slal}{\mathfrak{sl}}
\newcommand{\gl}{\mathfrak{gl}}
\begin{document}
\begin{titlepage}
\vskip 0.2cm
\begin{center}
{\Large {\bf Supergravity Black Holes and Billiards 
and
}}\\
{\Large {\bf
 Liouville integrable structure of dual Borel algebras$^\dagger$
}}\\[1cm]
{\large Pietro Fr\'e$^{\tt a}$  and Alexander S. Sorin$^{\tt b}$}
{}~\\
\quad \\
{{\em $^{a}$ Italian Embassy in the Russian Federation, \\
Denezhny Pereulok, 5, 121002 Moscow, Russia\\
{\tt  pietro.fre@esteri.it}\\
{\tt and}
\\
Dipartimento di Fisica Teorica, Universit\'a di Torino,}}
\\
{{\em $\&$ INFN - Sezione di Torino}}\\
{\em via P. Giuria 1, I-10125 Torino, Italy}~\quad\\
{\tt   fre@to.infn.it}
{}~\\
\quad \\
{{\em $^{\tt b}$ Bogoliubov Laboratory of Theoretical Physics,}}\\
{{\em Joint Institute for Nuclear Research,}}\\
{\em 141980 Dubna, Moscow Region, Russia}~\quad\\
{\tt sorin@theor.jinr.ru}
\quad \\
\end{center}
~{}
\begin{abstract}
In this paper we show that the supergravity equations describing
both cosmic billiards and a large class of black-holes are,
generically, both Liouville integrable as a consequence of the
same universal mechanism. This latter is provided by the Liouville
integrable Poissonian structure existing on the dual Borel algebra
$\mathbb{B}_\mathbb{N}$ of the simple Lie algebra
$A_\mathrm{N-1}$. As a by product we derive the explicit
integration algorithm associated with all symmetric spaces
$\mathrm{U/H}^\star$ relevant to the description of time-like and
space-like $p$-branes. The most important consequence of our
approach is the explicit construction of a complete set of
conserved involutive hamiltonians $\{ \mathfrak{h}_\alpha \}$
that are responsible for integrability and provide a new tool to
classify flows and orbits. We believe that these
 will prove a very important  new tool in
the analysis of supergravity black holes and billiards.
\end{abstract}
\vfill \vspace{2mm} \vfill \hrule width 3.cm {\footnotesize $^
\dagger $ This work is supported in part  by the Italian Ministry
of University (MIUR) under contracts PRIN 2007-024045. Furthermore
the work of A.S. was partially supported by the RFBR Grants No.
09-02-12417-$\mathrm{ofi\_m}$ , 09-02-00725-a, 09-02-91349-$\mathrm{NNIO\_a}$;
DFG grant No 436 RUS/113/669, and the Heisenberg-Landau
Program.}
\end{titlepage}
\tableofcontents
\section{Introduction}
\label{intro} Explicit supergravity solutions of pure and
matter-coupled supergravity in diverse dimensions play an
important role in the study of solitonic and instantonic states of
superstring theory, in particular $p$-brane states
\cite{Polchinski:1995mt,Gutperle:2002ai}.
\par
Indeed one large, diversified and important class of supergravity
solutions\footnote{For a review and for a large set of references
see for instance \cite{Stelle:2001ms}.} is provided by the
$p$-brane ones that are divided in two subclasses:
\begin{itemize}
  \item The space-like $p$-brane solutions that have an Euclidian world-volume and are time-dependent, all fields being functions of the time parameter $t$.
  \item The time-like $p$-brane solutions that have a Minkowskian world volume and are stationary, the
  fields depending on another parameter $t$, typically measuring the distance from the brane.
\end{itemize}
A view-point independently  introduced in \cite{Gal'tsov:1998yu}
and \cite{noiconsasha}, and systematically developed in
\cite{Weylnashpaper}, \cite{noipaintgroup}, \cite{noiKacmodpaper},
\cite{sahaedio}, \cite{Fre':2007hd}, identified, at least for
space-branes, the field equations of supergravity corresponding to
such solutions with the geodesic equations on the corresponding
moduli space that is mostly a homogeneous space and most
frequently also a symmetric space $\mathrm{U/H}$. This
identification allowed the in-depth study of supergravity cosmic
billiards \cite{ivshuk}, \cite{cosmicbilliardliterature1},
\cite{cosmicbilliardliterature2},
\cite{cosmicbilliardliterature3}, \cite{cosmicbilliardliterature4}
and lead to the discovery of their complete integrability
\cite{sahaedio}, \cite{Fre':2007hd}.
\par
It was an idea already circulating for some time in the community that also the construction of time-like $p$-branes,
in particular rotational symmetric black-hole solutions, could be reduced to the problem of geodesic motion on appropriate moduli spaces that would, this time, be Lorentzian
rather Euclidean coset manifolds $\mathrm{U/H}^\star$. This idea found a precise formulation in the recent publication
\cite{Bergshoeff:2008be}.
\par
In connection with these applications, the question of integrability of the differential systems of equations describing geodesic motion on homogeneous spaces and in particular on symmetric non-compact cosets $\mathrm{U/H}$, acquires particular relevance. As we emphasize in section \ref{sezionedue}, this question  is intimately related with the issue of \textit{normed solvable Lie algebras}, namely
solvable Lie algebras $\mathcal{S}$ equipped with a non-degenerate norm $< \, ,\,>$, which is positive definite in the space-brane (=billiard) case and indefinite in the time-brane (=black hole) case.
\par
In this paper, by performing a change of logical reference frame that replaces the route from geometry to Lie algebra into the opposite one and by gluing together pieces of mathematical knowledge dispersed in the literature, we show that:
\begin{enumerate}
  \item The integrability of all the various homogeneous models, both Euclidian and Lorentzian follows from the Liouville integrability of a universal parent model,
  associated with the Borel subalgebra $\mathbb{B}_\mathrm{N}$ of the $A_{\mathrm{N-1}}$ Lie algebra. The integrability of the parent extends to its
      children algebras $\mathcal{S}$ if the always existing embedding $\mathcal{S}\hookrightarrow \mathbb{B}_\mathrm{N}$ is adequate.
\item Liouville integrability of $\mathbb{B}_\mathrm{N}$ is an intrinsic property
of this algebra which allows to construct an adequate number of
universal hamiltonians $\left\{\mathfrak{h}_\alpha\right\}$ in
involution.
  \item The norm $< \, ,\,>$ on any  solvable Lie algebra $\mathcal{S}$
is not an independent external datum, rather it is intrinsically
defined by the restriction to $\mathcal{S}$ of the unique
quadratic hamiltonian $\mathfrak{h}_0$ on $\mathbb{B}_\mathrm{N}$,
once the embedding $\mathcal{S}\hookrightarrow
\mathbb{B}_\mathrm{N}$ has been defined.
  \item All symmetric coset models $\mathrm{U/H}^\star$ defined as follows have integrable
   geodesic equations. The Lie algebra $\mathbb{U}$ of the numerator is non-compact and the Lie algebra
  $\mathbb{H}^\star$ of the denominator is any of the real sections contained in $\mathbb{U}$ of the complexification $\mathbb{H}_\mathbb{C}$ of $\mathbb{H} \subset \mathbb{U}$, the former being the maximal compact subalgebra of the latter.
  \item The explicit integration algorithm has a universal form.
\end{enumerate}
Based on our new view-point we also present a new algorithmic approach to the study of the (eventual) integrability of
homogeneous normal spaces that are not symmetric spaces, leaving however the actual use of such an algorithm to future publications.
\section{A new view-point from old results}
\label{sezionedue} In this section we first summarize the basic
facts about the Riemannian or pseudo-Riemannian structures that
can be defined on a \textit{normed} solvable Lie algebra. Our goal
is that of reviewing the construction of the so named Nomizu
connection and of its associated  geodesic differential equations.
The reason is that we aim at a Copernican Revolution. In this
context, the classical route was from Riemannian geometry to Lie
algebra theory since the problems that motivated the consideration
of such mathematical structures were differential geometric in
nature: in particular the geometry of scalar manifolds appearing
in supergravity theories. It was very helpful and rewarding to
find a translation vocabulary that allowed the reformulation of
Riemannian geometry into a purely Lie algebraic setup. Yet, in
relation with integrability, this classical route obscures one
relevant fact:  integrability (when it exists) is an a priori
intrinsic property of the solvable Lie algebra. This property is
intelligently, yet secretly, utilized by the (pseudo)-Riemannian
structures. Hence, following our announced Copernican Revolution,
we aim at reverting the route, going from solvable Lie algebra
theory to (pseudo)-Riemanian geometry, rather than vice-versa.
This change of reference frame will prove very helpful in view of
old mathematical results, that were a little bit known in the
literature on non-linear science \cite{arh1}, \cite{deift1},
\cite{kodama1} but which had so far completely escaped
consideration in the current supergravity and superstring
literature.
\par
In the next subsection we prepare our Copernican Revolution with a
short review of the Ptolemaic system.
\subsection{The Ptolemaic system: Nomizu connection on a normal metric
solvable Lie algebra}
\par
Let us consider a solvable Lie algebra $\mathcal{S}$. For instance
$\mathcal{S}$ can be the Borel subalgebra of a complex semi-simple
Lie algebra $\mathbb{G}_{\mathbb{C}}$, namely\footnote{We recall
that given the Cartan-Weyl basis of a complex simple Lie algebra
$\mathbb{G}_{\mathbb{C}}$, its Borel subalgebra
$\mathrm{B}(\mathbb{G}_{\mathbb{C}})$ is defined as the solvable
algebra spanned by all the Cartan generators $\mathcal{H}_i$ and
by all the step operators $E^\alpha$ associated with all positive
roots $\alpha >0$.}:
\begin{equation}\label{borellusdefi}
    \mathcal{S}\,=\, \mathrm{B}(\mathbb{G}_{\mathbb{C}}) \, \equiv \, \mbox{span} \,\left \{ \mathcal{H}_i \, , \, E^\alpha \right \}
\end{equation}
or it can be the solvable Lie algebra canonically associated with
the pair made by a real form $\mathbb{G}_{\mathbb{R}} $ of
$\mathbb{G}_{\mathbb{C}}$ and by its maximal compact subalgebra
$\mathbb{H}_c \, \subset \, \mathbb{G}_{\mathbb{R}}$ \footnote{We
recall that the systematic construction of the solvable Lie
algebras associated with non-compact symmetric spaces, pioneered
in \cite{primisolvi} and then extensively developed in the
literature, has played a very important role in addressing,
solving and systematizing a large number of supergravity problems
associated with black-hole solutions \cite{otherBHpape},
\cite{noie7blackholes}, \cite{mario1}, \cite{mario2}, with
supergravity gaugings \cite{gaugedsugrapot}, \cite{myparis} and
later also with the issue of cosmic billiards introduced in
\cite{ivshuk}, \cite{cosmicbilliardliterature1},
\cite{cosmicbilliardliterature2},
\cite{cosmicbilliardliterature3}, \cite{cosmicbilliardliterature4}
and developed with the systematic help of the solvable Lie algebra
representation of supergravity scalar manifolds in
\cite{noiconsasha}, \cite{noiKacmodpaper}, \cite{Weylnashpaper},
\cite{noipaintgroup}, \cite{sahaedio}, \cite{Fre':2007hd},
\cite{Fre:2008zd}.}:
\begin{equation}\label{solvGH}
    \mathcal{S} \, = \, \Solv \left(
    \mathbb{G}_{\mathbb{R}}/\mathbb{H}_c\right)~.
\end{equation}
Other relevant choices of the solvable Lie algebra $\mathcal{S}$
can be made among those associated with the classification of
homogeneous special geometries that appear in the coupling to
matter of supergravity theories with eight supercharges in $D=5$,
$D=4$ and $D=3$ dimensions\footnote{We recall that the
classification of special homogeneous manifolds began with the
mathematical work of Alekseveesky in 1975 who posed himself the
problem of constructing all quaternionic K\"ahler manifolds with a
transitive solvable group of isometries \cite{Alekseevsky1975} and
then was completed and inserted into the $c$-map framework
\cite{Cecotti:1988ad} of supergravity with the work of de Wit et.
al. in \cite{deWit:1992wf}. Further studies continued in
\cite{Cortes} and for a complete recent discussion of the topic
and for all relevant further references we refer the reader to
\cite{contoine}.}:
\begin{equation}\label{specgeo}
    \mathcal{S} \quad \mbox{such that} \quad \left( \exp[\mathcal{S}] \, , \, <\, , \, >\right) \, = \, \mbox{\it SUGRA special Riemannian manifolds~.}
\end{equation}
The above writing refers to the main point of the Ptolemaic system
namely to the notion of \textit{normed metric solvable Lie
Algebras}. Following the original viewpoint of Alekseevsky we say
that a Riemannian manifold $\left ( \mathcal{M},g \right)$ is
\textit{normal} if it admits a completely solvable Lie group $\exp
[\Solv_{\mathcal{M}}]$ of isometries that acts on the manifold in
a simply transitive manner (i.e. for every 2 points in the
manifold there is one and only one group element connecting them).
The group $\exp[\Solv_{\mathcal{M}}]$ is then generated by a
so-called \textit{normal metric Lie algebra}, that is a completely
solvable Lie algebra $\Solv_{\mathcal{M}}$ endowed with an
Euclidean, positive definite, symmetric form $< \, , \, >$ . The
main tool to classify and study the normal homogeneous spaces is
provided by the theorem \cite{BorelTits}, \cite{Helgason} that
states that if a Riemannian manifold $\left ( \mathcal{M},g
\right)$ admits a transitive normal solvable group of isometries
$\exp[\Solv_{\mathcal{M}}]$, then it is metrically equivalent to
this solvable group manifold
 \begin{eqnarray}\label{identifico}
\mathcal{M} & \simeq & \exp \left[ \Solv_{\mathcal{M}}\,
\right]~,\nonumber  \\ g\mid_{e \in \mathcal{M}} & = & <,> \,
 \end{eqnarray}
where $<,>$ is the
Euclidean metric defined on the normal solvable Lie algebra
$\Solv_{\mathcal{M}}$.
\par
The conjecture of Alekseevsky was just restricted  to quaternionic
K\"ahler manifolds and  implied that any such manifold
$\mathcal{M}$ that was also homogeneous and  of negative Ricci
curvature  should   be normal, in the sense over mentioned, namely
a transitive solvable group of isometries $\exp \left
[\Solv_{\mathcal{M}}\right ]$ should exist, that could  be
identified with the manifold itself. Note that the actual group of
isometries $\mathrm{U}$ of $\mathcal{M}$ could be much larger than
the solvable group,
\begin{equation}\label{fattispecie}
    \mathrm{U} \, \supset \, \exp \left
    [\Solv_{\mathcal{M}}\right]~,
\end{equation}
as it is for instance the case for all symmetric spaces
\begin{equation}\label{pallus}
    \mathcal{M} \, = \, \frac{\mathrm{U}}{\mathrm{H}}
\end{equation}
yet the solvable normed Lie algebra $\left (\Solv_{\mathcal{M}}\,
, \, <\, ,\, >\right)$ had to exist. The problem of classifying
the considered manifolds was turned in this way into the problem
of classifying the \textit{normal metric solvable Lie algebras}
$\left(\mathcal{S}, <\, ,\, >\right)$. Note that in Alekseevsky's
case the symmetric form $<\, ,\, >$ was not only required to be
positive definite but also quaternionic K\"ahler. Alekseevsky's
conjecture  actually applies to more general homogeneous
Riemannian manifolds than the quaternionic ones: for instance it
applies to all those endowed with a special K\"ahler geometry or
with a real special one as the classification of de Wit et. al.
\cite{deWit:1992wf} demonstrated. It also applies to the symmetric
spaces appearing in the scalar sector of extended supergravities
with more than eight supercharges. For all these manifolds there
exists the corresponding normal metric  algebra
$\left(\mathcal{S}, <\, ,\, >\right)$, in other words they are
\textit{normal}. This happens because they are Einstein manifolds
of negative Ricci curvature and, although we are not aware of any
formal mathematical statement in this direction, one might make
the
\begin{congettura}\label{congetto}
$<<$ Every homogeneous Einstein manifold $\mathcal{M}$ of negative
Ricci curvature is normal, namely there exists a normal metric
solvable Lie algebra $\left(\mathcal{S}, <\, ,\, >\right)$ such
that identifying $\mathcal{S}$ with $\Solv_\mathcal{M}$ eq.
(\ref{identifico}) applies. $>>$
\end{congettura}
\par
Proving such a conjecture amounts to proving that for every
homogeneous Einstein manifold the group of isometries
$\mathrm{U}$, which by hypothesis of homogeneity exists and has a
transitive action on the manifold, admits a solvable
\textit{simply transitive subgroup}\footnote{Simply transitive
means that each group element has no fixed points.}
$\exp[\mathcal{S}] \subset \mathrm{U}$. If this is true, in view
of the already mentioned theorem the rest follows. The key
assumption is the negative Ricci curvature. Manifolds of positive
Ricci curvature, which are typically compact, are excluded. All
compact symmetric spaces are indeed counterexamples. For
$\mathrm{U/H}$ compact there is no transitive solvable subgroup of
$\mathrm{U}$.
\par
The recollection of these well known facts was done in order to
emphasize the following point. In the Ptolemaic system that starts
from Riemannian geometry and arrives at solvable Lie algebras
$\mathcal{S}$, this latter emerges in conjunction with a well
defined metric form $< \, ,\, >$ defined over it. For instance if
we focus on Borel solvable algebras
$\mathbb{B}(\mathbb{G}_{\mathbb{C}})$, they are endowed with the
following canonical metric:
\begin{eqnarray}\label{canonicB}
     < \mathcal{H}_i\, ,\, \mathcal{H}_j> & = &  2 \, \delta_{ij}~, \nonumber\\
     < \mathcal{H}_i\, ,\, E^\alpha> & = &  0 ~,\nonumber\\
     < E^\alpha\, ,\, E^\beta> & = &  \delta_{\alpha\beta}
\end{eqnarray}
whose normalization is absolute if the generators
$\{\mathcal{H}_i,E^\alpha,E^{-\alpha}\}$ of the Weyl-Cartan basis
for $\mathbb{G}_{\mathbb{C}}$ have the standard normalization
\begin{eqnarray}
  \left[ \mathcal{H}_i\, ,\, \mathcal{H}_j \right ] &=& 0~, \nonumber\\
  \left[ \mathcal{H}_i\, ,\, E^\alpha \right ] &=& \alpha_i \, E^\alpha ~,\nonumber\\
  \left[ E^\alpha \, ,\, E^\beta \right ]&=& N(\alpha,\beta) \,
  E^{\alpha + \beta} \quad \mbox{if $\alpha+\beta$ is a root} \quad ; \quad |N(\alpha,\beta)| = 1 ~,\nonumber \\
  \left[ E^\alpha \, ,\, E^{-\alpha} \right ] &=& 2 \, \alpha \, \cdot \, \mathcal{H}~, \nonumber\\
  \alpha \cdot \beta &=& \sum_{i=1}^{rank} \, \alpha_i \, \beta_i  \quad \Rightarrow \quad \{\mathcal{H}_i\} =
  \mbox{orthonormalized basis}~.
\end{eqnarray}
The metric (\ref{canonicB}) is singled out by the relation of the
Borel algebra $\mathbb{B}(\mathbb{G}_{\mathbb{C}})$ with one
specific Riemannian Einstein manifold of negative Ricci curvature.
This latter is
\begin{equation}\label{maxisplit}
    \mathcal{M} \, = \, \frac{\mathrm{G}_{\mathrm{ms}}}{\mathrm{H}_\mathrm{c}}
\end{equation}
where $\mathrm{G}_{\mathrm{ms}}$ denotes the group generated by
the unique real section $\mathbb{G}_{\mathrm{ms}}$ of the complex
Lie algebra $\mathbb{G}_{\mathbb{C}}$ which is maximally split
(or, equivalently maximally non-compact) and
$\mathrm{H}_\mathrm{c}$ denotes the unique maximally compact
subgroup of $\mathrm{G}_{\mathrm{ms}}$. It turns out that
$\mathbb{B}(\mathbb{G}_{\mathbb{C}})$ is just the solvable Lie
algebra of this Riemannian manifold $\mathcal{M}$
\begin{equation}\label{agnisco}
    \mathbb{B}(\mathbb{G}_{\mathbb{C}}) \, = \, \Solv_\mathcal{M}
\end{equation}
and the metric $< \, ,\, >$ (\ref{canonicB}) on the solvable Lie
algebra $\Solv_\mathcal{M}$ is that induced by the unique Einstein
Riemannian metric on the corresponding coset
$\frac{\mathrm{G}_{\mathrm{ms}}}{\mathrm{H}_\mathrm{c}}$
(\ref{maxisplit}).
\par
Similarly it happens in all other constructions of the Ptolemaic
system. The requirements imposed on the final Riemannian Einstein
manifold one wants to construct predetermine the metric $< \, ,\,
>$ on the solvable Lie algebra.
\par
Once the metric form is given, the construction of geometry and of
the associated geodesic equations follow uniquely. The issue is
just that of calculating the Levi--Civita connection of the metric
$g$ induced on the manifold by the form $< \, ,\, >$ defined on
the solvable Lie algebra. One way of describing this Levi--Civita
connection is by means of the so called \textit{Nomizu operator}
acting on  $\mathcal{S}$. The latter is defined as follows:
\begin{eqnarray}\label{nomizuAb}
 \mathbb{L}& : & \mathcal{S} \, \otimes\,
\mathcal{S} \rightarrow  \mathcal{S}, \\
 \forall X, Y, Z \in
\mathcal{S} & : & 2<\mathbb{L}_XY,Z> = <[X,Y],Z> - <X,[Y,Z]> - <Y,[X,Z]>.\nonumber
\end{eqnarray}
The \textit{Riemann curvature operator} on $\mathcal{S}$ can be expressed as
\begin{equation} \Riem(X,Y) =
[\mathbb{L}_X,\mathbb{L}_Y] - \mathbb{L}_{[X,Y]}~.
\end{equation}
If we introduce a basis of generators $\{T_A\}$ for $\mathcal{S}$
and the corresponding structure constants defined by
\begin{equation}\label{strutteconste}
    \left [ T_A \, , \, T_B \right ] \, = \, f_{AB}^{\phantom{AB}C} \, T_C
\end{equation}
together with the metric tensor:
\begin{equation}\label{Gciccius}
    < T_A ,T_B> \, = \, g_{AB}
\end{equation}
the connection defined by eq.(\ref{nomizuAb}) leads to the following  connection coefficients:
\begin{eqnarray}
  \mathbb{L}_A T_B\, &=& \Gamma_{AB}^C \, T_C \nonumber\\
  \Gamma_{AB}^C &=&  f_{AB}^{\phantom{AB}C} \, - \, g_{AD} g^{CE}\, f_{BE}^{\phantom{AB}D}
  - \, g_{BD} g^{CE}\, f_{AE}^{\phantom{AB}D}
\end{eqnarray}
which are constant numbers.
\par
Equivalently, we can define the Levi--Civita--Nomizu connection
starting from the dual description of the solvable Lie algebra in
terms of Maurer--Cartan equations. Let $e^A$ be a basis of
Maurer--Cartan forms dual to the generators $T_B$, namely
$e^A(T_B) \,= \delta^A_B$. We have
\begin{equation}\label{MCeque}
    d \, e^C \, = \, \ft 12 \, f_{AB}^{\phantom{AB}C} \, e^A \, \wedge e^B
\end{equation}
Interpreting $e^A$ as the vielbein over the solvable group
manifold we write the vanishing torsion equation
\begin{equation}\label{zerotorsion}
    0 \, = \, de^A \, + \, \omega^{AB} \, \wedge \, e^C \, g_{BC}
\end{equation}
where $\omega^{AB} \, = - \, \omega^{BA}$ is the standard
$\so(n)$--Lie algebra valued spin connection
($n=\mbox{dim}(\mathcal{S})$~). The relation between the two
descriptions is immediate:
\begin{equation}\label{fortissima}
    \omega^{AB} \, = \, \Gamma_{DE}^A \, g^{DB} \, e^E
\end{equation}
where the tensor $\Gamma_{DE}^A \, g^{DB}$ is automatically antisymmetric in force of its definition.
\par
Given the connection coefficients the differential geodesic equations can be immediately written. In the chosen basis
the tangent vector to the geodesic is described by $n$ fields $Y^A(t)$ which depend on the affine parameter $t$ along
the curve. The geodesic equation is given by the following first order differential system:
\begin{equation}\label{geoeque}
    \frac{d}{dt} \,Y^{A} \, + \, \Gamma^{A}_{BC} \, Y^B \, Y^C \, = \,
    0~.
\end{equation}
The above equation contains two data:
\begin{description}
  \item[1)] the structure constants of the solvable Lie algebra
  $f_{AB}^{\phantom{AB}C}$,
  \item[2)] the metric tensor $g_{AB}$.
\end{description}
As we emphasized the second datum, namely the metric, comes down, in the Ptolemaic system from the geometric
interpretation of the differential system (\ref{geoeque}) as geodesic equations on a Riemannian Einstein manifold
and it is the metric of that manifold what eventually predetermines $g_{AB}$.
\par
Let us now implement our Copernican Revolution and let us forget
for a moment about the Riemannian structure. The system
(\ref{geoeque}) is just a non linear differential system defined
over the dual $\mathcal{S}^*$ of a Solvable Lie algebra
$\mathcal{S}$. How could we directly derive such a differential
system and may be established its Liouville integrability from the
very structure of $\mathcal{S}$? Is there a way of deriving the
metric tensor implicitly contained in (\ref{geoeque}) from the Lie
algebra $\mathcal{S}$? There is.
\subsection{The Copernican Revolution: Poissonian structure of $\mathcal{S}$ and Liouville integrability}
\label{borellus} Given the solvable Lie algebra $\mathcal{S}$   we
can consider the co-adjoint orbits of the corresponding solvable
group $ \exp[\mathcal{S}]$. In simple language this amounts to
consider functions defined over the dual Lie algebra
$\mathcal{S}^*$. An element of $\mathcal{S}^\star$ is a linear
functional
\begin{equation}\label{phifuntor}
    \mathcal{L} \quad : \quad \mathcal{S} \, \rightarrow \, \mathbb{C}  \,\, \mathrm{or} \,\,\mathbb{R}
\end{equation}
and using a general theorem in linear algebra the dual  of a
finite dimensional vector space is isomorphic to the same space.
In practice, given a basis $T_A$ of $\mathcal{S}$ we immediately
obtain the dual basis $\mathcal{L}^A$ by defining
\begin{equation}\label{pudduduale}
    \mathcal{L}^A (T_B) \, = \, \delta^A_B~.
\end{equation}
Any dual Lie algebra element $\mathfrak{Y}\,\in
\,\mathcal{S}^\star$ can be written as a linear combination of the
dual generators
\begin{equation}\label{fraccoY}
    \mathfrak{Y} \, = \, Y_A \, \mathcal{L}^A
\end{equation}
and the most general function $\phi$ on $\mathcal{S}^\star$ is
actually a function of the $n$ coordinates $Y_A$. Given any two
such functions $\phi_1$ and $\phi_2$ we can define their
Lie--Poisson bracket in the following manner:
\begin{eqnarray}
\{\mathcal{\phi}_1, ~\mathcal{\phi}_2\} &=&
f_{AB}{}^C\,Y_C\,\frac{\partial \mathcal{\phi}_1}{\partial Y_A}\,
\frac{\partial \mathcal{\phi}_2}{\partial Y_B}\,~.\label{Poisson}
\end{eqnarray}
In this way the space of co-adjoint orbits becomes a Poisson manifold, independently from the existence of any metric
$< \, , \, >$ on $\mathcal{S}$. Then one is allowed to consider evolution equations of the following form:
\begin{eqnarray}
\frac{d}{dt}{Y}_A +\{{Y}_A, ~\mathcal{H}\}
&=&0\,\label{EulerHamilton}
\end{eqnarray}
where $\mathcal{H}=\mathcal{H}(Y)$ is some function on the dual
Lie algebra $\mathcal{S}^\star$ that we can regard as the
Hamiltonian.
\par
The question is whether the geodesic equations (\ref{geoeque}) can
be put in the hamiltonian form (\ref{EulerHamilton}), namely,
whether there exists a hamiltonian function, necessarily
quadratic, which reproduces the Levi--Civita--Nomizu connection.
The answer is obviously yes. It suffices to write
\begin{equation}
\mathcal{H} \,\equiv\, \frac{1}{2}
~{Y}_A\,{Y}_B\,g^{AB}\label{Hamil}
\end{equation}
and identify the variables $Y^A$ and $Y_B$ through the relation
\begin{equation}\label{sugiu}
    Y_A \, = \, g_{AB} \, Y^B~.
\end{equation}
Both in eq. (\ref{Hamil}) and in eq.(\ref{sugiu}) there appears the metric $g_{AB}$ defined
by the non degenerate normal form $< \, , \,>$. Hence it may seem that we made no
real progress and we simply rewrote the same equations in a different style. The notion of the metric tensor
is still essential and it looks external to the pure Lie algebraic structure. It is not so, as it appears from
the following argument.
\subsubsection{Liouville integrability}
The key point we would like to emphasize is that the definition of
the Lie--Poisson bracket (\ref{Poisson}) depends only on the
structure constants of the algebra $\mathcal{S}$ and nothing else,
so it is intrinsic to the algebra. Let us next recall the notion
of Liouville integrability.
\begin{definizione}
A symplectic manifold of dimension $2m$ endowed with a
Lie--Poisson bracket $\{\, , \,\}$ is Liouville integrable if
there exists $m$ functionally independent functions
$\Phi_i(\mathcal{Y})$ of its $2m$ coordinates $\mathcal{Y}_I$ that
are in involution. Namely we must have:
\begin{equation}\label{LiouvilleInt}
    \forall \, i,j \, = \, 1, \dots \, m \quad : \quad \left \{ \Phi_i \, , \, \Phi_j \right \} \, = \,0
\end{equation}
and
\begin{equation}\label{rankus}
    \mbox{$\mathrm{rank}$} \left(\, \frac{\partial{\Phi_i}}{\partial \mathcal{Y}_J} \right )\, = \,
    m~.
\end{equation}
\end{definizione}
When these conditions are fulfilled any of the functions $\Phi_i$
can be chosen as the hamiltonian $\mathcal{H}$ and the
corresponding Euler equations
\begin{equation}\label{EulerLag}
  0 \, = \,  \frac{d}{dt} \, \mathcal{Y}_I \, + \, \left \{ \mathcal{Y}_I \, , \, \mathcal{H}\, \right \}
\end{equation}
are completely integrable, since they admit $m$ first integrals of
the motion $\Phi_i(\mathcal{Y})$.
\par
Let us now envisage the following:
\subsubsection{Scenario}
\label{teatro}
Given the solvable Lie algebra $\mathcal{S}$, whose dimension we denote
\begin{equation}\label{nforte}
    {d}_\mathcal{S} \, \equiv \, \mathrm{dim} \, \mathcal{S}~,
\end{equation}
imagine that with respect to its intrinsic Lie--Poissonian
structure (\ref{Poisson}) there exist $p_\mathcal{S}$  functions
$\mathfrak{h}_\alpha(Y)$ of the coordinates
$Y_A~(A=1,...,{d}_\mathcal{S})$ on $\mathcal{S}^\star$ that are in
involution
\begin{equation}\label{involuzione}
    \left \{ \mathfrak{h}_\alpha \, , \, \mathfrak{h}_\beta \,\right \} \, = \, 0 \quad (\alpha,\beta\, = \,
 \, 1, \dots \,, p_\mathcal{S} )
\end{equation}
with the integer $p_\mathcal{S}$ lying in the range:
\begin{equation}\label{rangepS}
    d_\mathcal{S} \, \ge \, p_\mathcal{S} \, \ge \left[ \, \frac{d_\mathcal{S}}{2} \,\right]
\end{equation}
and suppose furthermore that, having defined the integer
\begin{equation}\label{numercasimir}
        c \, \equiv \,  2\,p_\mathcal{S} - \, d_\mathcal{S}~,
\end{equation}
 precisely $c$ of the functions $\mathfrak{h}_\alpha(Y)$, which,
for this reason we rename  $\mathcal{C}_\ell(Y)$
($\ell=1,\dots\,c$), are \textit{Casimirs}, namely they commute
with all the coordinates:
 \begin{equation}\label{casimirdefi}
    \left \{ Y_A \, , \, \mathcal{C}_\ell \,\right \} \, = \, 0  \quad ; \quad
     (\ell\, =\, 1,\dots,c \,\, ; \,\, A\, = \, 1, \dots \, d_\mathcal{S}
     )~.
 \end{equation}
Under these hypotheses, by rearranging the set of involutive functions
as it follows:
\begin{equation}\label{rearrangio}
    \left\{ \mathfrak{h}_\alpha \right \} \, = \, \left\{ \underbrace{\Phi_i(Y)}_{i=1,\dots,m_\mathcal{S}} \, , \,
    \underbrace{C_\ell(Y)}_{\ell=1,\dots ,c} \right \}
\end{equation}
we obtain the following situation. The \textit{level surfaces}
$\mathfrak{P}_{r_1,\dots,r_c}$ defined by setting the $c$ Casimirs
to fixed values
\begin{equation}\label{livelloni}
    \mathfrak{y} \, \in \, \mathfrak{P}_{r_1,\dots,r_c} \,\Leftrightarrow \, \left \{
    \begin{array}{ccc}
      \mathcal{C}_1 (\mathfrak{y})& = & r_1 \, \in \, \mathbb{R} \,  \, \left (\mathrm{or} \, ~\mathbb{C}\right )\\
      \mathcal{C}_2 (\mathfrak{y})& = & r_2 \, \in \, \mathbb{R} \,  \, \left (\mathrm{or} \, ~\mathbb{C}\right ) \\
      \dots & \dots & \dots \\
      \mathcal{C}_c (\mathfrak{y})& = & r_c \, \in \, \mathbb{R} \,  \, \left (\mathrm{or} \, ~\mathbb{C}\right ) \\
    \end{array}\right.
\end{equation}
are by definition manifolds of even dimension:
\begin{eqnarray}\label{simpletlivelli}
    \mathrm{dim} \, \mathfrak{P}_{r_1,\dots,r_c} & = &  2 \, m_\mathcal{S}\nonumber\\
    m_\mathcal{S} & = & \frac{d_\mathcal{S} - c}{2} \, = \, d_\mathcal{S} \, - \, p_\mathcal{S}
\end{eqnarray}
and on these manifolds there exist exactly $m_\mathcal{S}$
functions in involution, namely the $\Phi_i(Y)$ of
eq.(\ref{rearrangio}). It follows that each of these manifolds
acquires a symplectic hamiltonian structure. Naming
$\mathcal{Y}_I$ the $2\, m_\mathcal{S}$ coordinates on
$\mathfrak{P}_{r_1,\dots,r_c}$ and choosing as hamiltonian any
linear combination of the pull-backs on
$\mathfrak{P}_{r_1,\dots,r_c}$ of the functions $\Phi_i(Y)$
\begin{equation}\label{hamilcombo}
    \mathcal{H}(\mathcal{Y}) \, = \, \mbox{\rm pull-back on $\mathfrak{P}_{r_1,\dots,r_c}$ of } \, a^i \,
    \Phi_i(Y)
\end{equation}
the corresponding Euler equation differential system
\begin{equation}\label{eulercombo}
 0 \, = \,    \frac{d}{dt} \, \mathcal{Y}_I \, + \, \left \{ \mathcal{Y}_I \, , \, \mathcal{H} \right \}
\end{equation}
is by construction Liouville integrable. Indeed the pull-backs of
the $m_\mathcal{S}$ functions $\Phi_i(Y)$ provide the necessary
first integrals of motion.
\subsubsection{Realization of this scenario on $\mathbb{B}( A_{\mathrm{N-1}})$}
The above described scenario is precisely realized in a very
notable case, namely that of the Borel subalgebra
$\mathbb{B}({A}_{{\mathrm{N-1}}}) \subset {A}_{{\mathrm{N-1}}}$.
As it is well known the simple Lie algebra $A_{\mathrm{N-1}}$,
identified by the following Dynkin diagram:
\begin{center}
\begin{picture}(100,100)
\put (-70,85){$A_{\mathrm{N-1}}$}
\put (-20,90){\circle {10}}
\put (-23,75){$\alpha_1$}
\put (-15,90){\line (1,0){20}}
\put (10,90){\circle {10}}
\put (7,75){$\alpha_2$}
\put (15,90){\line (1,0){20}}
\put (40,90){\circle {10}}
\put (37,75){$\alpha_3$}
\put (47,90){$\dots$}
\put (70,90){\circle {10}}
\put (67,75){$\alpha_{N-3}$}
\put (75,90){\line (1,0){20}}
\put (100,90){\circle {10}}
\put (97,75){$\alpha_{N-2}$}
\put (105,90){\line (1,0){20}}
\put (130,90){\circle {10}}
\put (127,75){$\alpha_{N-1}$}
\end{picture}
\end{center}
is the abstract form of the Lie algebra
$\slal(\mathrm{N};\mathbb{C})$ of complex traceless matrices in
dimension $\mathrm{N}$. The corresponding Borel subalgebra
$\mathbb{B}_\mathrm{N} \, \equiv
\,\mathbb{B}({A}_{{\mathrm{N-1}}})$ is simply given by  the subset
of all upper triangular traceless matrices. It is more convenient
to relax the condition on the trace and consider the Borel
subalgebra $\widehat{\mathbb{B}}_\mathrm{N} \, \equiv \,
{\mathbb{B}}\left(\gl(\mathrm{N};\mathbb{C})\right)$ which is
simply made by all upper triangular matrices. Reduction to
${\mathbb{B}}_\mathrm{N}$ will be performed by putting one of the
Casimirs to the null value. Hence we define
\begin{equation}
\label{Borelliano} \gl (\mathrm{N};\mathbb{C}) \, \supset \,
\widehat{\mathbb{B}}_\mathrm{N} \, \ni \, \mathfrak{b} \, = \,
\left(
  \begin{array}{cccccc}
    \star & \star & \star & \star & \dots & \star \\
    0 & \star & \star & \star & \dots & \star \\
    0 & 0 & \star & \star & \dots & \star \\
    0 & 0 & 0 & \star & \dots & \star \\
    \dots & \dots & \dots & \dots & \dots & \dots \\
    0 & 0 & 0 & 0 & 0  & \star \\
  \end{array}
\right)~.
\end{equation}
The dimension of this solvable algebra is easily computed
\begin{equation}\label{dimenBN}
    \mathrm{dim} \,\widehat{\mathbb{B}}_\mathrm{N} \, \equiv \, d_\mathrm{N} \, = \,
    \frac{\mathrm{N}(\mathrm{N}+1)}{2}~.
\end{equation}
It was demonstrated by Arhangel'skii in \cite{arh1} that this
Poissonian manifold is Liouville integrable according to the
scheme described in the previous subsection and the procedure to
construct the required set of functions in involution was
described in \cite{arh1,deift1}. Let us review these results. We
distinguish the two cases where $\mathrm{N} \, = \, 2 \nu $ is
even and where $\mathrm{N} \, = \, 2 \nu+1$ is odd. The number
$p_\mathrm{N}$ of functions in involution, the number of Casimirs
and the ensuing even dimension of the orbits is displayed below:
{\small
\begin{equation}\label{numerelli}
    \begin{array}{|c|c|c|c|c|c|c}
    \hline
       \null & d \, = \, \mathrm{dim} \,\widehat{\mathbb{B}} & p \, \equiv \, \# \,\mbox{funct. in inv.} & c \, \equiv \, \# \,\mbox{of Casim.} & 2m \, = \, \mbox{dim of orbits}\\
       \hline
       \widehat{\mathbb{B}}_{2\nu} & \nu(2\nu+1) & \nu^2 + \nu & \nu & 2\nu^2 \\
       \widehat{\mathbb{B}}_{2\nu+1}& (\nu+1)(2\nu+1) & \nu^2+2\nu+1 & \nu +1 & 2(\nu^2 + \nu) \\
       \hline
     \end{array}
\end{equation}
}
 For the reader's convenience in Table \ref{troika} we tabulated the first
instances of such numbers.
\begin{table}[!tb]
\begin{center}
$
    \begin{array}{|c|c|c|c|c|}
    \hline
      \mathrm{N} & d_\mathrm{N} & p_\mathrm{N} & c_\mathrm{N} & 2m \\
      \hline
     2 & 3 & 2 & 1 & 2 \\
 3 & 6 & 4 & 2 & 4 \\
 4 & 10 & 6 & 2 & 8 \\
 5 & 15 & 9 & 3 & 12 \\
 6 & 21 & 12 & 3 & 18 \\
 7 & 28 & 16 & 4 & 24 \\
 8 & 36 & 20 & 4 & 32 \\
 9 & 45 & 25 & 5 & 40 \\
 10 & 55 & 30 & 5 & 50 \\
 11 & 66 & 36 & 6 & 60 \\
 12 & 78 & 42 & 6 & 72 \\
 13 & 91 & 49 & 7 & 84 \\
 14 & 105 & 56 & 7 & 98 \\
 15 & 120 & 64 & 8 & 112 \\
 16 & 136 & 72 & 8 & 128\\
      \hline
    \end{array}
$
\end{center}
\caption{\it In this table we give the dimensions of the Borel
algebras $\mathbb{B}(\gl(\mathrm{N}))$, the corresponding number
 of functions in involutions, the corresponding number of Casimirs and the ensuing dimensions
of the orbits for the first sixteen values of $\mathrm{N}$.\label{troika}}
\end{table}

\subsubsection{Involutive hamiltonians}
How are the functions $\mathfrak{h}_{\alpha}(Y)$ explicitly
constructed?

First let us observe that a simple set of coordinates on the Borel
subalgebra ${\mathbb{B}}(\gl(\mathrm{N}))$ is simply given by the
entries of the upper triangular matrix $\mathfrak{b}$ mentioned in
eq.(\ref{Borelliano}). Secondly, for reasons that will become
clear later on, let us consider an $\mathrm{N}\times \mathrm{N}$
matrix $L$ which satisfies the condition
\begin{eqnarray}
(L~\eta)^{T} = L~\eta
 \label{LaxProp}
\end{eqnarray}
where
\begin{eqnarray}
\eta = \diag\left(-1,+1,...,-1,+1,+1,...+1\right)~, \quad p \leq q \quad ; \quad p+q \, = \, \mathrm{N}
 \label{metric}
\end{eqnarray}
is any of the available choices of $(p,q)$ signatures in dimension
$\mathrm{N}$. Irrespectively of the choice of $\eta$, the number
of parameters contained in  a matrix  $L$ satisfying
(\ref{LaxProp}) is always equal to $\mathrm{N(N+1)}/2$ which is
the dimension of the Borel algebra $\mathbb{B}(\gl(\mathrm{N}))$.
Indeed there is a simple one-to-one map from the space of upper
triangular matrices to the space of matrices satisfying
eq.(\ref{LaxProp}), which reads as follows:
\begin{eqnarray}
\label{mappino}
  \mathfrak{b} & \mapsto & L \, = \, \mathfrak{b} \, + \, \eta \, \mathfrak{b}^T \, \eta ~,\nonumber\\
  L &\mapsto & \mathfrak{b} \, = \, L_{>}\, -\, \frac{1}{2}~\diag{L}~.
\end{eqnarray}
In the above equation we have used the following convention. For
any matrix $M$, we denote by $M_{>}$ its upper triangular part and
by $M_{<}$ its lower triangular one including the diagonals.
\par
Hence the space of matrices fulfilling eq.(\ref{LaxProp}) provides
a coordinate basis for the Borel algebra. Thanks to the simple
relation (\ref{mappino}), the use of different $\eta$ tensors just
amounts to a \textit{linear coordinate transformation} on
$\widehat{\mathbb{B}}_\mathrm{N}$, in other words to a change of
basis for the generators of the solvable Lie algebra. What is then
the relevance of considering such different choices of $\eta$
rather than focusing on a single conventional one? The answer is
elementary. Each different $\eta$ prepares a basis well adapted to
the decomposition of the Lie algebra $\gl(\mathrm{N})$ with
respect to its subalgebra $\so(p,q)$. Such a decomposition is
necessary in order to study the geometry and the associated
geodesic equations of the coset manifold
\begin{equation}\label{cosettusino}
   \mathcal{M}_{p,q} \, = \,
   \frac{\mathrm{GL(p+q)}}{\mathrm{SO(p,q)}}~.
\end{equation}
All these $(p,q)$ systems are integrable since all the manifolds
$\mathcal{M}_{p,q}$ are metrically equivalent to the same solvable
manifold $\exp [\widehat{\mathbb{B}}_{p+q}]$, equipped however
with a different normal form $< \, ,\,>_{p,q}$, which is positive
definite only for $\{p=0,~q=\mathrm{N}\}$. The Liouville
integrability of all these systems has the same common root,
namely the existence of the $p_\mathrm{N}$ independent functions
in involution that now we display and which is an intrinsic
property of the Borel algebra $\widehat{\mathbb{B}}_{p+q}$.
\par
The algorithm of constructing these functions, originally derived
for $\{p=0,~q=\mathrm{N}\}$ in \cite{arh1} and \cite{deift1},
being generalized to the case under consideration, is the
following.
\par
Starting from the parameterization of
$\widehat{\mathbb{B}}_{\mathrm{N}}$ by means of the matrix $L$
(\ref{mappino}) fulfilling eq.(\ref{LaxProp})
 the complete set of $p_\mathrm{N}$ functions $\mathfrak{h}_\alpha $ that are involutive with
respect to the Lie--Poisson bracket (\ref{Poisson}) is enumerated
by an ordered pair of indices
\begin{equation}\label{orderedpair}
    \alpha \, = \, (a,b)
\end{equation}
where:
\begin{eqnarray}
  a &=&  0,...,\left[\frac{\mathrm{N}}{2}\right]~,\nonumber\\
  b &=& 1,..., \mathrm{N}-2a ~.
\end{eqnarray}
The functions $\mathfrak{h}_{ab}$
can be iteratively derived from the following relation:
\begin{eqnarray}
\label{tuttehamile} &&\det\left\{(L-\lambda)_{ij}:~a+1\leq i \leq
\mathrm{N}, ~ 1\leq j \leq
\mathrm{N}-a\right\}\nonumber\\&&=\mathcal{E}_{a 0
}\left(\lambda^{\mathrm{N}-2a} +\sum_{b=1}^{\mathrm{N}-2a}
\mathfrak{h}_{a b}~\lambda^{N-2a-b}\right), \quad a =
0,...,\left[\frac{\mathrm{N}}{2} \right]\label{Hamiltonians}
\end{eqnarray}
where, by definition, $\mathcal{E}_{a 0}$ is the coefficient of the power
$\lambda^{\mathrm{N}-2a}$.
\par
As we know from the previous discussions $c_\mathrm{N}$ of these \textit{generalized hamiltonians} are actually Casimirs.
\par
For example, for the case $\mathrm{N}=3$ that we study in detail in the next sections
the functions $\mathfrak{h}_{02}$, $\mathfrak{h}_{03}$ are pure hamiltonians while
$\mathfrak{h}_{01}$, $\mathfrak{h}_{11}$ are Casimir functions.
\subsection{Integrability and the metric $< \, ,\,>_{p,q}$}
Having established in purely algebraic intrinsic terms the
integrability of the differential system based on the solvable Lie
algebra $\mathcal{S} \, = \,\widehat{\mathbb{B}}_\mathrm{N}$,  we
can now answer the question about the origin of the metric form $<
\, ,\,>_{p,q}$ which, in the Ptolemaic system, leads to the very
same differential equations. The problem is very simply solved by
recalling the quadratic form (\ref{Hamil}) of the hamiltonian
which yields the geodesic equations associated with the
Levi--Civita--Nomizu connection. The  metric $g_{AB}$ is defined
by the coefficients appearing in general unique \textit{quadratic}
hamiltonian function $\mathcal{H} =\mathfrak{h}_{02}$. This
view-point is actually very efficient. Changing the metric
signature $\eta$ amounts, as we have emphasized, to a change of
basis on the solvable Lie algebra. Using all the time the same
formula (\ref{tuttehamile}) to calculate the hamiltonians
$\mathfrak{h}_\alpha$, but varying the choice of $\eta$ in the
definition of $L$ (\ref{mappino}), results in a change of the
coefficients in the quadratic hamiltonian
$\mathfrak{h}_{\alpha_0}$, namely in a change of metric $g_{AB}$
over the Lie algebra. All the various coset models with indefinite
signature are covered in this way by a unique algorithm.
\subsection{Integrability and solvable subalgebras}
In view of the above results on the integrability of all Borel algebras $\widehat{\mathbb{B}}_\mathrm{N}$ there arises a natural
question about the relation of this integrability with the possible integrability of its subalgebras. Indeed
in view of the theorem that states that every linear representation of a solvable algebra $\mathcal{S}$ admits
a basis where all the  elements $X\in \mathcal{S}$ are represented by an upper triangular matrix it follows that any solvable Lie algebra can be regarded as a subalgebra $\mathcal{S} \subset \widehat{\mathbb{B}}_\mathrm{N}$ for a suitable choice of $\mathrm{N}$.
\par
Hence let us consider a generic subalgebra
$\mathcal{S}\subset\widehat{\mathbb{B}}_\mathrm{N}$ and name
$\widetilde{\mathfrak{h}}_\alpha$ the pull-back on
$\mathcal{S}^\star$ of the $p_\mathrm{N}$ generalized hamiltonians
defined on $\widehat{\mathbb{B}}_\mathrm{N}$. In particular we are
just interested in the differential system constructed on
$\mathcal{S}^\star$ where the Hamiltonian
$\widetilde{\mathcal{H}}$ is the pull-back of the unique quadratic
hamiltonian $\mathcal{H}= \mathfrak{h}_{02}$ constructed on
$\widehat{\mathbb{B}}_\mathrm{N}$. From the mathematical point of
view there might be other choices but from the physical point of
view this is the only relevant one. So let us put
\begin{equation}\label{pullobacco}
    \widetilde{\mathcal{H}} \, = \, \widetilde{\mathfrak{h}}_{02}
\end{equation}
and let us divide the $\widehat{\mathbb{B}}_\mathrm{N}$
coordinates in two subsets, those along $\mathcal{S}$ and those
normal to $\mathcal{S}$
\begin{equation}\label{sdoppio}
    \{ Y_A \} \, = \, \{ \underbrace{X_i}_{\in \mathcal{S}} \, , \, \underbrace{W_\alpha}_{\notin \mathcal{S}}
    \}~.
\end{equation}
Next let us compare the evolution equations of the subalgebra
$\mathcal{S}$ with the pull-back on $\mathcal{S}$ of the
$\mathbb{B}_N$ equations. In view of our choice (\ref{pullobacco})
we have
\begin{equation}\label{starro}
    \widetilde{\mathcal{H}}(X) \, = \, \mathcal{H}(X,W)|_{W=0} \, \equiv
    \,\mathcal{H}(X,0)~.
\end{equation}
The $\mathcal{S}$-equations are simply read
\begin{eqnarray}\label{Seque}
    \frac{d}{dt}{X}_i & = & \, - \,\{ X_i \, , \, \widetilde{\mathcal{H}}\} \nonumber\\
    & = & \, - \, f_{ij}^{\phantom{ij}k} X_k
    ~\partial^j\widetilde{\mathcal{H}}~.
\end{eqnarray}
while the pull-back on $\mathcal{S}$ of the
$\widehat{\mathbb{B}}_\mathrm{N}$ equations is
\begin{eqnarray}\label{Seque2}
    \frac{d}{dt}{X}_i & = & \, - \,\{ X_i \, , \, \mathcal{H}\}|_{W=0} \nonumber\\
    & = & \, - \, f_{ij}^{\phantom{ij}k} X_k \,\partial^j\widetilde{\mathcal{H}} \,  -  \,
    f_{i\alpha}^{\phantom{ij}k} X_k \,\partial^\alpha\mathcal{H}|_{W=0} ~,\, \, \label{Xpro}\\
    \frac{d}{dt}{W}_\alpha & = & \, - \,\{ W_\alpha \, , \, \mathcal{H}\}|_{W=0} \nonumber\\
    & = & \, - \, f_{\alpha \beta}^{\phantom{\alpha\beta}k}  X_k \, \partial^\beta{\mathcal{H}}|_{W=0} \, -  \,
    f_{\alpha i}^{\phantom{ij}k}  X_k \,\partial^i\widetilde{\mathcal{H}}~. \, \label{Wpro}
\end{eqnarray}
The conditions necessary for the system (\ref{Xpro},\ref{Wpro}) to reduce consistently to the system (\ref{Seque}) are the following ones:
\begin{eqnarray}
  \partial^\beta\mathcal{H}|_{W=0} & = & 0~,\\
  f_{\alpha i}^{\phantom{ij}k} &=& 0~.
\end{eqnarray}
The first condition is satisfied if there are no mixed
coefficients in the metric implicitly defined by the hamiltonian,
namely if
\begin{equation}\label{pallo}
    g^{i\alpha} \, =\, 0~.
\end{equation}
The second condition implies that the decomposition of the Borel
algebra $\widehat{\mathbb{B}}_\mathrm{N}$ with respect to its
solvable subalgebra $\mathcal{S}$ should be reductive. Altogether
the conditions for the consistent reduction of the system
(\ref{Xpro},\ref{Wpro}) to the system (\ref{Seque}) can be written
as follows:
\begin{eqnarray}
  \widehat{\mathbb{B}}_\mathrm{N} &=& \mathcal{S} \, \oplus \, \mathcal{S}_\perp~, \label{splitto}\\
  < \mathcal{S} \, , \, \mathcal{S}_\perp > & = & 0~, \label{ortoreduce}\\
  \left[\mathcal{S} \, , \, \mathcal{S} \right] & \subset & \mathcal{S}~, \label{subalgo}\\
  \left[\mathcal{S} \, , \, \mathcal{S}_\perp \right] & \subset & \mathcal{S}_\perp~.\label{reduzio}
\end{eqnarray}
Under the hypotheses (\ref{splitto}--\ref{reduzio}) a geodesic on
the manifold $(\exp[S],~\widetilde{< \, , \, >})$ satisfies the
geodesic equations on the embedding manifold
$(\exp[\widehat{\mathbb{B}}_\mathrm{N}],~< \, , \, >)$. The latter
is a symmetric space, namely
$\mathrm{GL(N,\mathbb{R})/SO(p,N-p)}$, so that
$(\exp[S],~\widetilde{< \, , \, >})$ is revealed to be a
\textit{geodesically complete submanifold} of a symmetric space.
In force of a general theorem \cite{Helgason} this implies that
also $(\exp[S],~\widetilde{< \, , \, >})$ is a \textit{symmetric
space}.
\par
Hence the exceptional cases of homogeneous solvable manifolds
$(\exp[\mathcal{S}],~\widetilde{< \, , \, >})$ that are not
symmetric spaces (compare with the classification of homogeneous
special geometries, \cite{deWit:1992wf}) are based on solvable Lie
algebras $\mathcal{S}$ whose embedding in
$\widehat{\mathbb{B}}_\mathrm{N}$ is not reductive and violates
eq.s (\ref{splitto}--\ref{reduzio}). In order to establish their
Liouville integrability one has to study the pull-back on
$\mathcal{S}$ of the $p_\mathrm{N}$ involutive hamiltonians
$\widetilde{\mathfrak{h}}_\alpha$ (see Table \ref{troika}) and
ascertain how many of them remain in involution among themselves
and with the unique hamiltonian
$\widetilde{\mathcal{H}}=\widetilde{\mathfrak{h}}_{02}$. Such a
study involves several complicacies and it is postponed to a
subsequent publication. For the case of symmetric spaces, instead,
the proper reduction of the hamiltonians and the consequent
integrability is guaranteed a priori from the proper reduction of
the Lax representation that we discuss in the next section.
\section{Triangular embedding in
$\mathrm{SL(N;\mathbb{R}) / SO(p,N-p;\mathbb{R})}$ and
integrability of the Lorentzian cosets
$\mathrm{U}/\mathrm{H}^\star$} As already recalled in the
introduction it was realized in recent years that the field
equations of supergravity describing either
\begin{enumerate}
   \item[1)] space-like $p$-branes as the cosmic billiards, or
   \item[2)] time-like $p$-branes as several rotational invariant black-holes in $D=4$ and more general solitonic branes
  in diverse dimensions
\end{enumerate}
reduce to geodesic equations on coset manifolds of the type
\begin{equation}\label{UoverH}
    \mathcal{M} \, = \, \frac{\mathrm{U}}{\mathrm{H}} \quad or \quad \mathcal{M}^\star \, = \, \frac{\mathrm{U}}{\mathrm{H}^\star}
\end{equation}
where $\mathrm{\mathrm{U}} = \exp[\mathbb{U}]$ is the group manifold generated by the duality algebra $\mathbb{U}$ relevant to the
considered supergravity model in the considered dimensions. As a rule without exceptions $\mathbb{U}$ is always
some \textit{non-compact real form} of a complex Lie algebra $\mathbb{U}_\mathbb{C}$. The Lie algebra $\mathbb{H}$ of the subgroup $\mathrm{H}$ is instead the unique maximal compact subalgebra of $\mathbb{U}$.
The coset $\mathcal{M}$ corresponds to the case of
space-like $p$-branes, in particular cosmic billiards. In the second coset $\mathcal{M}^\star$, pertaining to
the case of time-like branes, the Lie algebra $\mathbb{H}^\star$ of $\mathrm{H}^\star$ is another non-compact real section of
the complexification $\mathbb{H}_\mathbb{C}$, while $\mathbb{U}$ remains the same. It happens indeed the following situation always occurs: the real duality algebra $\mathbb{U}$ admits as proper subalgebras  a few different instances
of real sections of $\mathbb{H}_\mathrm{C}$ which necessarily include the maximally compact one $\mathbb{H}$.
\par
We recall few illustrative examples. Consider the case of maximal supersymmetry with 32 supercharges. The duality algebra in $D=4$ is $\mathbb{U}=\mathrm{E_{7(7)}}$ whose maximal compact subalgebra is $\mathbb{H} = \su(8) \subset \mathrm{E_{7(7)}}$.
The complexification of this is $\mathbb{H}_\mathbb{C} \, = \, \slal(8,\mathbb{C})$. Aside from $\su(8)$ the other real
sections of $\slal(8,\mathbb{C})$ which are subalgebras of $\mathrm{E_{7(7)}}$ are the following ones:
$\mathbb{H}^\star \, = \,\slal(8,\mathbb{R}) \, \subset \, \mathrm{E_{7(7)}}$ and $\mathbb{H}^\star \, = \,\su^\star(8) \,
\subset \, \mathrm{E_{7(7)}}$. For the same number of supercharges the duality algebra in $D=5$ is $\mathbb{U}=\mathrm{E_{6(6)}}$. Here we have $\mathbb{H}=\usp(8)$, $\mathbb{H}_\mathbb{C} \, = \, \sym(8,\mathbb{C})$,
$\mathbb{H}^\star \, = \, \usp(4,4)$ or $\mathbb{H}^\star \, = \, \sym(8,\mathbb{R})$.
\par
Consider instead the case of $1/2$ maximal supersymmetry, namely supergravity with $16$ supercharges. In $D=3$
the theory with $6+n$ supervector multiplets has the following duality algebra $\mathrm{U}=\so(8,8+n)$.
The corresponding maximal compact subalgebra is $\mathbb{H}=\so(8)\times \so(8+n)$ and the available $\mathbb{H}^\star$ are:
$\mathbb{H}^\star \, = \, \so(8-p,p) \times \so(8+n-p,p)$ (p=1,2,3,4).
\par
For all these cases we can make the following statement.
\par
\begin{statement}\label{statamento}
$<<$ Let $\mathrm{N}$ be the real dimension of the fundamental
representation of $\mathbb{U}$. For each choice of $\mathbb{H}$ or
$\mathbb{H}^\star$ there exist a suitable integer $p\le
\left[\frac{\mathrm{N}}{2} \right]$ and a diagonal metric
\begin{equation}
\eta =
\diag(\underbrace{-1,+1,...,-1,+1}_{2p},\underbrace{+1,+1,...+1}_{\mathrm{N}-2p})~,
 \label{etametric}
\end{equation}
such that we have a canonical embedding
\begin{eqnarray}
 \mathbb{ U} & \hookrightarrow & \slal(\mathrm{N};\mathbb{R})~,\nonumber\\
\mathbb{ U} \, \supset \, \mathbb{H}^\star & \hookrightarrow &
\so\mathrm{(p,N-p;\mathbb{R})} \, \subset \,
\slal(\mathrm{N};\mathbb{R})~. \label{triaembed}
\end{eqnarray}
This embedding is determined by the choice of the basis where $\Solv\left(\mathrm{U/H^\star}
\right)$ is made by upper triangular matrices. In the same basis the
elements of $\mathbb{K}$ are $\eta$-symmetric matrices while those of $\mathbb{H}^\star$ are
$\eta$-antisymmetric ones, namely:
\begin{eqnarray}\label{etasym}
    \forall \, \mathrm{K}\, \in \, \mathbb{K} & : & \quad \eta \, \mathrm{K}^T \, = \,  \mathrm{K}^T \, \eta~, \nonumber\\
    \forall \, \mathrm{H}\, \in \, \mathbb{H}^\star & : & \quad \eta \, \mathrm{H}^T \, = \, - \,  \mathrm{H}^T \,
    \eta~.
\end{eqnarray}
$>>$
\end{statement}
Just as in the case of cosmic billiards the embedding of the
Riemannian coset $\mathrm{U/H}$ into the universal covering coset
$\mathrm{SL(N;\mathbb{R})/SO(N;\mathbb{R})}$ provided the key to
obtain an explicit integration algorithm for the associated first
order geodesic equations, in the same way the embedding
(\ref{triaembed}) of the pseudo-Riemannian $\mathrm{U/H^\star}$
into $\mathrm{SL(N;\mathbb{R})/SO(p,N-p;\mathbb{R})}$ provides the
key to extend the same integration algorithm also to this
indefinite metric symmetric spaces. Indeed that algorithm is
defined for $\mathrm{SL(N;\mathbb{R})/SO(p,N-p;\mathbb{R})}$ and
it has the property that if initial data are defined in a
submanifold $\mathrm{U/H^\star}$ where $\mathrm{U} \, \subset \,
\mathrm{SL(N;\mathbb{R})}$ and $\mathrm{H^\star} \, \subset \,
\mathrm{SO(p,N-p;\mathbb{R})}$, then the entire \textit{time flow}
occurs in the same submanifold. This property is a simple
consequence of the following argument. Let $L \in \mathbb{K}$ be
an $\eta$-symmetric matrix satisfying property (\ref{LaxProp})
which lies in the subspace $\mathbb{K}$ corresponding to the
symmetric decomposition $\mathbb{U}=\mathbb{H}^\star\oplus
\mathbb{K}$ of the Lie subalgebra $\mathbb{U} \subset
\gl(\mathrm{N};\mathbb{R})$. By construction the corresponding
connection $ W\equiv L_{>} - L_{<}$ lies in $\mathbb{H}^\star$,
namely we have $W\in \mathbb{H}^\star$. It follows that $[W, \, \,
L] \in \mathbb{K}$. Hence if $L(0) \in \mathbb{K}$, the evolution
equation $\dot{L}\, + \, [W, \, \, L]=0$ never brings $L(t)$ out
of that space.
\par
Hence the embedding
(\ref{triaembed}) suffices to define explicit integration formulae for
all supergravity time-like $p$-branes based on pseudo-Riemannian symmetric spaces.
\par
Let us review the steps of the procedure.
\begin{enumerate}
  \item First one defines a coset representative for $\mathrm{U/H^\star}$  just as in the Riemannian case, namely:
\begin{equation}
  \mathbb{L}\left( \phi\right)  \, = \, \prod_{I=m}^{I=1} \,
  \exp \left [\varphi_I \,
  E^{\alpha_I}\right] \, \exp \left[h_i \mathcal{H}^i\right]
\label{cosettorepresentat}
\end{equation}
where the roots pertaining to the solvable Lie algebra are ordered in
ascending order of height ($\alpha_I \le \alpha_J$ if $I < J$), $\mathcal{H}^i$ denote the non compact Cartan generators
and the product of matrix exponentials appearing in (\ref{cosettorepresentat}) goes from the highest
on the left, to lowest root on the right. In this
way the parameters $\left \{\phi\right\} \, \equiv \, \left \{\varphi_I \, , \, h_i \right\} $ have a precise and uniquely
defined correspondence with the fields of supergravity by means of dimensional oxidation \cite{noiconsasha,Weylnashpaper}. Thanks to the mapping (\ref{mappino}), the same upper triangular matrix $\mathbb{L}\left( \phi\right)$ can be regarded as a coset representative for $\mathrm{U/H}$ or for $\mathrm{U/H^\star}$.
  \item Restricting all the fields $\phi$ of supergravity to pure
  \textit{time dependence}\footnote{In the case of time like $p$-branes such
as black-holes \textit{time} is actually the radial coordinate.}
$\phi \, = \, \phi(t)$, the coset representative becomes also a
function of
  \textit{time} $\mathbb{L}\left(\phi(t) \right) = \mathbb{L}(t)$ and we define the Lax operator
  $L(t)$ and the connection $W(t)$ as follows:
\begin{eqnarray}
  L(t) & = & \sum_{i} \, \mbox{Tr} \left(\mathbb{L}^{-1}
  \frac{d}{dt}\mathbb{L} \, \mathrm{K}_i\right) \mathrm{K}_i~,
  \nonumber\\
    W(t) & = & \sum_{\ell} \, \mbox{Tr} \left(\mathbb{L}^{-1}
  \frac{d}{dt}\mathbb{L} \, \mathrm{H}_\ell\right)\,
  \mathrm{H}_\ell
\label{postayanna}
\end{eqnarray}
where $\mathrm{K}_i$ and $\mathrm{H}_\ell$ denote an orthonormal
basis of generators for $\mathbb{K}$ and $\mathbb{H}^\star$,
respectively
\begin{eqnarray}\label{segnoni}
    \mbox{Tr}\left( \mathrm{K}_i\, , \, \mathrm{K}_j\right ) & = & \mbox{diag}\left(\underbrace{+,\dots
      ,\,+}_{n-r}\, , \underbrace{-,\,\dots \, , \, -}_{r}\right)~,\nonumber\\
    \mbox{Tr}\left( \mathrm{H}_i\, , \, \mathrm{H}_j\right ) & = & 2\, \mbox{diag}\left(\underbrace{+,\dots ,\,+}_{r}\,, \underbrace{-,\,\dots \, , \,-}_{m-r}\right)~,\nonumber\\
    \mbox{Tr}\left( \mathrm{K}_i\, , \, \mathrm{H}_j\right ) & = & 0~,\nonumber\\
    n & \equiv& \mbox{dim}\left ( \frac{\mathrm{U}}{\mathrm{H}^\star}\right) \, = \, \mbox{dim}\left ( \frac{\mathrm{U}}{\mathrm{H}}\right)~, \nonumber\\
    m & \equiv & \mbox{dim}\left ( \mathrm{H}^\star\right) \, = \, \mbox{dim}\left ( \mathrm{H}\right)~, \nonumber\\
    r & \equiv & \# \,\mbox{compact generators of $\mathbb{K}$} \, = \, \# \,\mbox{non-compact generators of $\mathbb{H}$}~.\nonumber\\
\end{eqnarray}
\item With these definitions the field equations of the supergravity time-like $p$-brane or black-hole,
which are just the geodesic equations for the manifold $\mathrm{U/H^\star}$ in
the solvable parametrization, reduce to the single matrix valued Lax
equation \cite{sahaedio}
\begin{eqnarray}
\label{Lax} \frac{d}{dt} L(t)+\left [W (t)\, , \, L(t)\right ] \,
= \, 0 \label{newLax}
\end{eqnarray}
which is the compatibility condition for the linear system
exhibiting the iso-spectral property of the Lax operator $L$
\begin{eqnarray}
\label{LaxLinSys1}
L~\Psi=\Psi~ \Lambda~,\\
\frac{d}{dt} \Psi = W~\Psi \label{LaxLinSys2}
\end{eqnarray}
where
\begin{eqnarray}
\Lambda = \diag\left(\lambda_1, ..., \lambda_{N}\right)
\label{lambda}
\end{eqnarray}
is the diagonal matrix of eigenvalues and $\Psi(t)$ is the
eigenmatrix.
\item If we are able to write the general integral of the Lax
equation, depending on $n=\mbox{dim}(\mathrm{U/H^\star})$ integration
constants, then comparison of the definition of the Lax operator
(\ref{postayanna},\ref{cosettorepresentat}) with its explicit form
in the integration reduces the differential equations of supergravity
to quadratures
\begin{equation}
  \frac{d}{dt} \phi(t) \, = \, F(t) \quad = \quad \mbox{known function of time.}
\label{quadrature}
\end{equation}
\end{enumerate}
\subsection{The integration algorithm for the Lax Equation}
Let us assume that we have explicitly constructed the embedding
(\ref{triaembed}). In this case, in the decomposition
\begin{equation}
  \mathbb{U} = \mathbb{K} \oplus \mathbb{H}^\star
\label{UKH}
\end{equation}
of the relevant Lie algebra $\mathbb{U}$, the matrices
representing the elements of $\mathbb{K}$ are all $\eta$-symmetric
while those representing the elements of $\mathbb{H}^\star$ are
all $\eta$-antisymmetric as we have already pointed out.
Furthermore the matrices representing  the solvable Lie algebra
$\Solv(\mathrm{U}/\mathrm{H}^\star)$ are all upper triangular.
These are the necessary and sufficient conditions to apply to the
relevant Lax equation (\ref{Lax}) the integration algorithm
originally described in \cite{kodama1} and reviewed in
\cite{sahaedio,Fre':2007hd}. The key point is that the connection
$W(t)$ appearing in eq.(\ref{newLax}) is related to the Lax
operator by means of the already recalled algebraic projection
operator as follows:
\begin{eqnarray}
 W=\Pi (L):=L_{>}-L_{<}~.
 \label{Lprojection}
\end{eqnarray}
The relation (\ref{Lprojection}) is nothing else but the statement
that the coset representative $\mathbb{L}(\phi)~(=L\,+\,W)$ from
which the Lax operator is extracted is taken in the solvable
parametrization.
\par
The only relevant new feature distinguishing the indefinite metric case from the definite one is the discussion of the
spectral types.
\subsubsection{Spectral types}
 From eqs.(\ref{LaxProp}) it follows that, generically, the Lax operator
$L(0)$ is not a symmetric matrix.  It is such only in the
Euclidean case ($p= 0$ and $q = \mathrm{N} $).  Therefore $L(0)$
eigenvalues are generically complex numbers. We will concentrate
on the case when $L(0)$ is a simple matrix, i.e. all its
eigenvalues are distinct. Then, in order for $L(0)$ to be real,
its eigenvalues have to group in $k$ complex conjugated pairs and
$\mathrm{N}-2k$ real eigenvalues, where $k$ is some fixed integer
in the range $0 \leq k \leq p$. Obviously, Lax matrices
corresponding to different values of $k$ can not be related by a
similarity transformation. Thus the integer $k ~$ parameterizes
inequivalent initial data $L_k(0)$ that we decide to name
\textit{spectral types}. For the metric (\ref{metric}) one can
choose a basis in the space of eigenvalues where their complex
conjugation properties become
\begin{eqnarray}
\lambda^{*}_{2\alpha-1} &=& \lambda_{2\alpha}~, \quad \alpha
=1,...,k~, \quad 0
\leq k~ \leq p~,\nonumber\\
\lambda^{*}_\alpha &=& \lambda_\alpha~, \quad
~\alpha=2k+1,...,\mathrm{N}~.
 \label{eigenvalues1}
\end{eqnarray}
Simple inspection of eqs. (\ref{eigenvalues1}) shows that the
corresponding matrix eigenvalues $\Lambda_k$ (\ref{lambda}) and
its complex conjugated matrix $\overline{\Lambda}_k$ are actually
related by a similarity transformation
\begin{eqnarray}
\overline{\Lambda}_k =  T_k~ \Lambda_k ~T^{-1}_k
 \label{eigenvalues2}
\end{eqnarray}
which will be useful in what follows. Here, the complex symmetric
$\mathrm{N}\times \mathrm{N}$--matrix $T_k$ has a very simple
block--diagonal structure
\begin{eqnarray}
T_k:=\diag\left(B,...,B, -\mathbf{1}_{\mathrm{N}-2k}\right)
 \label{Tmatrix}
\end{eqnarray}
and satisfies the properties
\begin{eqnarray}
\overline{T}_k ~T_k = \mathbf{1}_{\mathrm{N}}~, \quad T_k ~\eta~ T_k = \eta~.
 \label{propertiesT0}
\end{eqnarray}
The matrix $T_k$ comprises $k$ sub--blocks given by the $(2\times 2)$--matrix
$B$, defined below:
\begin{equation}
    B =  \left(
\begin{array}{ll}
 0 & {\rm i}  \\
 {\rm i} & 0
\end{array}
\right)~, \quad B^2  = - \mathbf{1}_{2}~, \label{block1}
\end{equation}
and the bottom $((\mathrm{N}-2k)\times (\mathrm{N}-2k))$--matrix
sub--block which is proportional to the $((\mathrm{N}-2k)\times
(\mathrm{N}-2k))$--unity matrix $\mathbf{1}_{\mathrm{N}-2k}$;
${\rm i}$ is the imaginary unity. The matrices $\Lambda_k$
($k=0,1,...,p$) satisfying eq. (\ref{eigenvalues2}) form the
initial data for the $L(0)$ eigenvalues.
\par
Now, let us turn to constructing the initial data for the
$L(0)$--eigenmatrix $\Psi(0)$ (\ref{LaxLinSys1}). If the Lax
eigenvalues are complex (i.e., $\overline{\Lambda}_k \neq
\Lambda_k$), then the eigenmatrix $\Psi(0)$ has to be complex as
well in order for the Lax matrix
\begin{eqnarray}
L_k(0)=\Psi_k(0)~\Lambda_k~ \Psi^{-1}_k(0)
 \label{L0}
\end{eqnarray}
to be real
\begin{eqnarray}
\overline{\Psi}_k(0)~\overline{\Lambda}_k~
{(\overline{\Psi}_k(0))}^{-1}=\Psi_k(0)~\Lambda_k ~\Psi^{-1}_k(0)~.
 \label{RealConstr}
\end{eqnarray}
Besides the constraint (\ref{RealConstr}), the complex eigenmatrix
$\Psi(0)$ which diagonalizes  the initial Lax operator $L(0)$
should satisfy one more constraint \cite{kodama1}
\begin{eqnarray}
\Psi^{T}_k(0) ~\eta ~\Psi_k(0)=\eta~,
 \label{property}
\end{eqnarray}
namely it should belong to the complexified group
$\mathrm{SO(p,N-p;\mathbb{C})}$. Let us introduce the new set of
block--diagonal complex, symmetric $\mathrm{N}\times
\mathrm{N}$--matrices
\begin{eqnarray}
\widehat{T}_k:=\diag\left(\frac{1}{\sqrt{2}}~(\mathbf{1}_{2}-B),...,
\frac{1}{\sqrt{2}}~(\mathbf{1}_{2}-B), \mathbf{1}_{N-2k}\right)
 \label{Thatmatrix}
\end{eqnarray}
that satisfy the following relations among themselves, with the
metric $\eta$ (\ref{metric}) and with the previously introduce set
$T_k$ (\ref{Tmatrix}):
\begin{eqnarray}
{\overline{\widehat{T}}}_k ~\widehat{T}_k =
\mathbf{1}_{\mathrm{N}}~, \quad \widehat{T}_k ~\eta~ \widehat{T}_k
= \eta~, \quad {\overline{\widehat{T}}}_k ~T_k =-\widehat{T}_k~.
 \label{propertiesT}
\end{eqnarray}
Using eqs.(\ref{eigenvalues2}) and (\ref{propertiesT})   by
straightforward  calculation we can verify that the complex eigenmatrix
\begin{eqnarray}
\Psi_k(0) := \mathcal{O}_0~\widehat{T}_k
 \label{Eigenmartix_final}
\end{eqnarray}
satisfies both constraints (\ref{RealConstr}) and (\ref{property})
if the introduced {\it real} matrix $\mathcal{O}_0$, entering into
eq. (\ref{Eigenmartix_final}), satisfies the pseudo--orthogonality
condition
\begin{eqnarray}
\mathcal{O}^{T}_0 ~\eta ~\mathcal{O}_0=\eta~,
 \label{orthogSOpq}
\end{eqnarray}
namely if $\mathcal{O}_0 \in \mathrm{SO(p,N-p;\mathbb{R})}$.
\par
We conclude that the initial data for the Lax operator are
represented in the following way:
\begin{eqnarray}
L_k(0)& = &\mathcal{O}_0~\widehat{T}_k~ \Lambda_k ~\widehat{T}^{-1}_k~
\mathcal{O}_0^{-1} \quad \mathrm{where}\nonumber\\
&& \mathcal{O}_0 \in \mathrm{SO(p,N-p;\mathbb{R})}~.
 \label{L0_last}
\end{eqnarray}
Hence the manifold of solutions of the Lax equation for the case $\mathrm{GL(N,\mathbb{R})/SO(p,N-p;\mathbb{R})}$, splits into $(\mathrm{p}+1)$ disconnected branches corresponding to the spectral types $k=0,1,..,p$.
\par
When we consider a subsystem $\mathrm{U}/\mathrm{H}^\star$, having
fixed $p$, the actual number of spectral types is determined by
considering which normal forms
\begin{equation}\label{normaliforme}
    \widehat{\Lambda}_{k} \, \equiv \, \widehat{T}_k~ \Lambda_k ~\widehat{T}^{-1}_k
\end{equation}
actually belong to the space $\mathbb{K}$. Indeed the spectral
type $k$ will be included in the integration algorithm if and only
if
\begin{equation}\label{pistacchio}
    \widehat{\Lambda}_{k} \, \in \, \mathbb{K}~.
\end{equation}
The above discussion of spectral types actually coincides with the
discussion of normal forms reported in
\cite{Bergshoeff:2008be}\footnote{We would like to thank our
friend and long term collaborator M. Trigiante who attracted our
attention to the constructions of paper
 \cite{Bergshoeff:2008be} when the results of the present paper
 were already in final form.}. In the same paper the normal form $\widehat{\Lambda}_{k} $ was group-theoretically
 interpreted as an element of the subspace $\mathbb{K}$ belonging to a subvector space of the form:
\begin{equation}\label{particolasanta}
     \widehat{\Lambda}_{k} \, \in \, \left(\frac{\slal(2;\mathbb{R})}{\so(1,1;\mathbb{R})} \right)^k \, \times \, \so(1,1;\mathbb{R})^{\mathrm{N}-k} \, \subset \,
     \mathbb{K}~.
\end{equation}
Furthermore in \cite{Bergshoeff:2008be} the number of actually available normal forms (spectral types in our
 nomenclature) was shown to admit the following group theoretical interpretation:
 \begin{equation}\label{numerspectyp}
    \# \, \mbox{\rm of spectral types} \, = \, \mbox{ \rm rank}\left( \frac{\mathrm{H}^\star}{\mathrm{H}_c}\right ) \, + \,1
 \end{equation}
where $\mathrm{H}_c \subset \mathrm{H}^\star$ is the maximal
compact subgroup of $H^\star$.
\subsubsection{The Kodama integration algorithm for $\frac{\mathrm{SL(N;\mathbb{R})}}{\mathrm{SO(p,N-p;\mathbb{R})}}$ revisited}
Having clarified the fundamental issue of spectral types let us
describe in full detail the adaptation of the Kodama integration
algorithm \cite{kodama1} to the indefinite metric case.
\par
Summarizing all our previous discussions the starting point
 of the algorithm is provided by the initial data listed below:
\begin{description}
\item[a)] The spectral type, codified by the choice of one of the $p+1$
matrices $\widehat{T}_{k}$ $(k=0,1,...p)$ (\ref{Thatmatrix}). At
fixed $k$ we set:
    \begin{equation}\label{spectraltype}
        \mathcal{T}\, := \, \widehat{T}_{k}~.
    \end{equation}
    \item[b)] The eigenvalue vector:
\begin{equation}\label{eigenvalvec}
    \overrightarrow{\lambda} \, = \,\left \{ \lambda_1,\lambda_2 ,\dots \, ,\lambda_{N-1}\,, \,-\sum_{i=1}^{N-1}\,
    \lambda_i\,\right\}~.
  \end{equation}
  If the spectral type is $k=0$ all the eigenvalues are real. If the spectral type is $k=r$ then the first
  $2r$ eigenvalues are arranged in complex conjugate pairs such that $\lambda_{2i}=\overline{\lambda}_{2i-1}$ while the remaining $\mathrm{N}-2r$ are real.
 \item[c)] The choice of an arbitrary element of the pseudo--rotational group $\mathcal{O}_0 \, \in \, \mathrm{SO(p,N-p;\mathbb{R})}$,  namely a real $\mathrm{N} \times \mathrm{N}$ matrix satisfying:
     \begin{equation}\label{Ocondo}
        \mathcal{O}_0^T \, \eta \, \mathcal{O}_0 \, = \, \eta~.
     \end{equation}
\end{description}
In terms of these initial data we define the following complex
matrices:
\begin{equation}\label{phipsinote}
    \Psi(0) \, = \, \mathcal{O}_0\, \mathcal{T} \, \in \,
    \mathrm{SO(p,N-p;\mathbb{C})}~.
\end{equation}
Next we construct a \textit{time-dependent}  $\mathrm{N} \times \mathrm{N}$ matrix $c(t)$ whose elements are defined by the following formula:
\begin{eqnarray}\label{cijt}
    c_{ij}(t) =  \sum_{k=1}^N \,\Psi_{ik}(0) \exp[-2\,\mathrm{Re}(\lambda_k) \,t]\,
    \left(\cos[-2\,\mathrm{Im}(\lambda_k)\,t]+{\rm i}\sin[-2\,\mathrm{Im}(\lambda_k)\,t]\right) \, \left(\Psi^{-1}(0)\right
    )_{kj}~,\nonumber\\
\end{eqnarray}
and  we introduce the $\mathrm{N}$-functions $D_\ell(t)$
($D_0(t):=0$) constructed as the determinants of the principal
diagonal $\ell \times \ell$ sub-matrices of $c(t)$, namely:
 \begin{eqnarray}
 D_\ell(t)=\mathrm{Det} \Biggr [ \Bigr ( c_{ij}(t) \Bigr )_{1\leq i,j \leq \ell} \Biggr
 ]~.
 \label{Ddefi}
\end{eqnarray}
Using these building blocks we construct a
\textit{time--dependent}, real, pseudo--rotation matrix
$\mathcal{O}(t)$ in the following way. First define a
\textit{time--dependent} complex matrix $\Psi(t)$ whose entries
are given by
\begin{eqnarray}
\Psi_{ij}(t)&=&\frac{\exp[-\mathrm{Re}\lambda_j
\,t](\cos[-\mathrm{Im}\lambda_j \,t]+{\rm i}
\sin[-\mathrm{Im}\lambda_j \,t])}{\sqrt{D_i(t)D_{i-1}(t)}} \,
\mathrm{Det} \, \left ( \begin{array}{cccc}
c_{11}&\dots &c_{1,i-1}& \Psi_{1j}(0)\\
\vdots&\ddots&\vdots&\vdots\\
c_{i1}&\dots &c_{i,i-1}& \Psi_{ij}(0)\\
\end{array}\right
)\nonumber\\
\label{Phidit}
\end{eqnarray}
and then we obtain the desired $\mathcal{O}(t)$ by setting
\begin{equation}\label{Otimedepe}
    \mathcal{O}(t) \, \equiv \, \Psi(t) \, \mathcal{T}^{-1}~.
\end{equation}
It is  a matter of direct verification that defined as above, the
evolving $\mathrm{SO(p,N-p)}$ group element $\mathcal{O}(t)$ is
indeed real at all instants of \textit{time}
\begin{equation}\label{realityOfO}
    \overline{\mathcal{O}}(t) \, = \, \mathcal{O}(t)~,
\end{equation}
thus ${\mathcal{O}}(t)\, \in \, \mathrm{SO(p,N-p;\mathbb{R})}$.
Finally the explicit form of the Lax operator solving Lax equation
with the chosen set of initial conditions is given by
\begin{eqnarray}
  L(t) &=& \mathcal{O}(t) \, \widehat{\Lambda} \, \mathcal{O}^{-1}(t)~, \\
  \widehat{\Lambda} &=& \mathcal{T} \, \Lambda \, \mathcal{T}^{-1}
\end{eqnarray}
where:
\begin{equation}\label{Lambdone}
    \Lambda \, = \, \left ( \begin{array}{c|c|c|c|c}
                                       \lambda_1 & 0 & \dots & 0 & 0 \\
                                       \hline
                                       0 & \lambda_2 & \dots & \dots & 0 \\
                                       \hline
                                       \dots & \dots & \dots & \dots & \dots \\
                                       \hline
                                       0 & 0 & \dots & \lambda_{N-1} & 0 \\
                                       \hline
                                       0 & 0 & \dots & 0 & -\sum_{i=1}^{N-1} \lambda_i
                                     \end{array} \right)~.
\end{equation}
At first sight the reader might consider baroque the substitution
\begin{equation}\label{subbarulla}
\exp[-2 \, \lambda_j \,  t] \, \mapsto \,  \exp[-2\,\mathrm{Re}(\lambda_k) \,t]\,
    \left(\cos[-2\,\mathrm{Im}(\lambda_k)\,t]+{\rm i}\sin[-2\,\mathrm{Im}(\lambda_k)\,t]\right)
\end{equation}
used both in eq. (\ref{cijt}) and eq. (\ref{Phidit}). Actually
such a substitution is very handy in order to verify the reality
of the solution  and also in order to understand its analytic
structure.  First of all it is fairly simple to check that the
matrix $c_{ij}(t)$ and hence all of its minors are real. Secondly
thanks to the same token one verifies that also $\mathcal{O}(t)$
is real and all of its entries are rational functions of
exponentials, cosines and sines or square-roots thereof. The
periodic trigonometric functions are absent for the spectral type
$k=0$ and appear only for $k\ge 1$. The appearance
 of these sines and cosines is the truely new feature due to the pseudo-Riemannian structure of the denominator
 group $\mathrm{H}^\star$.
\par
For all spectral types the evolution of the Lax operator is given
by a time dependent $\mathrm{SO(p,N-p;\mathbb{R})}$ similarity
transformation, starting from a normal form $\widehat{\Lambda}$,
but this normal form is diagonal only in the case of the spectral
type $k=0$ when all eigenvalues are real. In the other spectral
types the real normal form $\widehat{\Lambda}$ is non-diagonal. It
has  the structure discussed in the previous sections.
\section{The paradigmatic example: $\mathrm{SL(3;\mathbb{R})}/\mathrm{SO(1,2;\mathbb{R})}$ versus
$\mathrm{SL(3;\mathbb{R})}/\mathrm{SO(3;\mathbb{R})}$} In the
present section we illustrate the previously presented theory with
a simple, yet paradigmatic example, that of the cosets
\begin{equation}\label{M3p}
    \mathcal{M}_{p,3} \, = \, \frac{\mathrm{SL(3;\mathbb{R})}}{\mathrm{SO(p,3-p;\mathbb{R})}}
\end{equation}
where $p=1$ is the new case with respect to the well known example of $p=0$ (see for instance \cite{Fre':2007hd}).
 From the physical point of view the coset $\mathcal{M}_{0,3}$ provides the description of $D=5$ pure gravity reduced
three dimensions. Correspondingly the coset  $\mathcal{M}_{1,3}$
is just related with time-like brane solutions of $D=5$ pure
gravity. It can also be related to several other interesting brane
constructions, yet the viewpoint adopted in the present paper is
mathematically oriented. We just want to illustrate by means of
this examples the key points of the explained general
constructions. In particular we aim at illustrating the relation
between the choice of spectral type, the assignment of values to
the generalized hamiltonians in involution and the analytic
structure of the constructed integrals.
\par
Hence discarding the well known case $p=0$ for which we refer the
reader to \cite{Fre':2007hd}, we concentrate on the novel features
of the case $p=1$.
\subsection{$\so(1,2;\mathbb{R})$ decomposition of the $\mathrm{SL(3;\mathbb{R})}$ Lie algebra}
The starting point is to consider a basis of
$\mathrm{SL(3;\mathbb{R})}$ generators well adapted to the
Minkowskian coset
$\mathrm{SL(3,\mathbb{R})}/\mathrm{SO(1,2;\mathbb{R})}$ rather
than to the Euclidian coset
$\mathrm{SL(3,\mathbb{R})}/\mathrm{SO(3;\mathbb{R})}$. This is
easily constructed with the following procedure. First one
multiplies all generators adapted to the euclidian basis by the
Lorentzian metric tensor
\begin{equation}\label{etatensor}
    \eta \, = \, \left(
                          \begin{array}{ccc}
                            1 & 0 & 0 \\
                            0 & -1 & 0 \\
                            0 & 0 & -1 \\
                          \end{array}
                        \right)
\end{equation}
then one redefines the diagonal Cartan generators by subtracting
their trace and making them in this way traceless. The result is a
set of $8$ generators of the $\slal(3;\mathbb{R})$ Lie algebra
with the property that the last three close the Lie algebra of
$\so(1,2;\mathbb{R})$, while the first five span a basis of the
spin $s=2$ representation of the same. Namely we have:
\begin{eqnarray}\label{algener}
    \mathcal{T}_A & = & \left \{ \underbrace{\mathrm{\mathrm{K}_i}}_{i=1,\dots,5} \, , \,\underbrace{\mathrm{J_a}}_{a=1,\dots ,3} \right \}~,\nonumber\\
    0 & = & \eta \, \mathrm{J}_a^T \, + \mathrm{J}_a \, \eta~, \nonumber\\
    \left [ \, \mathrm{J}_a \, ,\, \mathrm{J}_b \right ] & = & \varepsilon_{abc} \, \eta^{cd} \,\mathrm{J}_d~, \nonumber\\
    \left [ \, \mathrm{J}_a \, ,\, \mathrm{K}_i \right ] & = &\mathrm{ R(J_a)}_{ij} \, \mathrm{K}_j
\end{eqnarray}
where $\varepsilon_{abc} $ is the standard Levi--Civita
antisymmetric symbol and $\mathrm{R}(\mathrm{J}_a)_{ij}$ denote
the $5\times 5$ matrices representing the $\so(1,2;\mathbb{R})$
generators in the spin $s=2$ case.
\par
Explicitly we have
\begin{equation}\label{sl3so12Kgen}
    \begin{array}{ccccccc}
       \mathrm{K}_1 & = & \left(
\begin{array}{lll}
 \frac{1}{\sqrt{2}} & 0 & 0 \\
 0 & -\frac{1}{\sqrt{2}} & 0 \\
 0 & 0 & 0
\end{array}
\right) & ; & \mathrm{K}_2 & = & \left(
\begin{array}{lll}
 -\frac{1}{\sqrt{6}} & 0 & 0
   \\
 0 & -\frac{1}{\sqrt{6}} & 0
   \\
 0 & 0 & \sqrt{\frac{2}{3}}
\end{array}
\right)~, \\
\null & \null & \null & \null &\null & \null &\null \\
       \mathrm{K}_3 & = & \left(
\begin{array}{lll}
 0 & -\frac{1}{\sqrt{2}} & 0
   \\
 \frac{1}{\sqrt{2}} & 0 & 0 \\
 0 & 0 & 0
\end{array}
\right) & ; & \mathrm{K}_4 & = & \left(
\begin{array}{lll}
 0 & 0 & 0 \\
 0 & 0 & -\frac{1}{\sqrt{2}}
   \\
 0 & -\frac{1}{\sqrt{2}} & 0
\end{array}
\right)~, \\
\null & \null & \null & \null &\null & \null &\null \\
       \mathrm{K}_5 & = & \left(
\begin{array}{lll}
 0 & 0 & -\frac{1}{\sqrt{2}}
   \\
 0 & 0 & 0 \\
 \frac{1}{\sqrt{2}} & 0 & 0
\end{array}
\right) & , & \null & \null & \null
     \end{array}
\end{equation}
and
\begin{equation}\label{sl3so12Jgen}
    \begin{array}{ccccccc}
       \mathrm{J}_1 & = & \left(
\begin{array}{lll}
 0 & -1 & 0 \\
 -1 & 0 & 0 \\
 0 & 0 & 0
\end{array}
\right) & ; & \mathrm{J}_2 & = & \left(
\begin{array}{lll}
 0 & 0 & 0 \\
 0 & 0 & -1 \\
 0 & 1 & 0
\end{array}
\right)~, \\
\null & \null & \null & \null &\null & \null &\null \\
       \mathrm{J}_3 & = & \left(
\begin{array}{lll}
 0 & 0 & -1 \\
 0 & 0 & 0 \\
 -1 & 0 & 0
\end{array}
\right) & . & \null & \null & \null\\
     \end{array}
\end{equation}
As one can easily note, among the coset generators three are non-compact, while two are compact.
\par
With reference to eq.(\ref{segnoni}) this means that in this case
we have
\begin{equation}\label{valoredir}
    r \, = \, 2~.
\end{equation}
Explicitly the non--compact coset generators are the two Cartans $\mathrm{K}_{1,2}$ and the off-diagonal generator $\mathrm{K}_4$. The two compact coset generators are instead $\mathrm{K}_{3,5}$. Similarly the three subalgebra generators are distributed
into the two non-compact ones $\mathrm{J}_{1,3}$ and the compact one $\mathrm{J}_2$. The explicit form of the $5\times 5$ matrices representing the generators $\mathrm{J}_{i}$ is the following one:
\begin{equation}\label{RRJrep}
    \begin{array}{ccccccc}
       \mathrm{R(J_1)}& = & \left(
\begin{array}{lllll}
 0 & 0 & -2 & 0 & 0 \\
 0 & 0 & 0 & 0 & 0 \\
 -2 & 0 & 0 & 0 & 0 \\
 0 & 0 & 0 & 0 & -1 \\
 0 & 0 & 0 & -1 & 0
\end{array}
\right) & ; & \mathrm{R(J_2)} & = &  \left(
\begin{array}{lllll}
 0 & 0 & 0 & 1 & 0 \\
 0 & 0 & 0 & \sqrt{3} & 0 \\
 0 & 0 & 0 & 0 & 1 \\
 -1 & -\sqrt{3} & 0 & 0 & 0 \\
 0 & 0 & -1 & 0 & 0
\end{array}
\right)~, \\
\null & \null & \null & \null &\null & \null &\null \\
       \mathrm{R(J_3)} & = & \left(
\begin{array}{lllll}
 0 & 0 & 0 & 0 & -1 \\
 0 & 0 & 0 & 0 & \sqrt{3} \\
 0 & 0 & 0 & -1 & 0 \\
 0 & 0 & -1 & 0 & 0 \\
 -1 & \sqrt{3} & 0 & 0 & 0
\end{array}
\right) & , & \null & \null & \null
     \end{array}
\end{equation}
and they are such that
\begin{equation}\label{Rmatrrole}
    \left [ \, \mathrm{J}_a \, ,\, \mathrm{K}_i \right ] \, = \,\mathrm{ R(J_a)_{ij}} \,
    \mathrm{K_j}~.
\end{equation}
\subsection{The Lorentzian Lax operator}
The next step is the definition of the Lax operator. According to our established conventions we define it as follows:
\begin{eqnarray}\label{LElax}
    L (t) & = & Y_1(t) \, \mathrm{K}_1 \, + \,  Y_2(t) \, \mathrm{K}_2 \, + \, \frac{1}{\sqrt{2}}\,\sum_{i=3}^5 \, Y_i(t) \mathrm{K}_i\nonumber\\
    \null & = & \left(
\begin{array}{lll}
 \frac{1}{\sqrt{2}}Y_1(t)-\frac{1}{\sqrt{6}}Y_2(t) & -\frac{1}{2}
   Y_3(t) & -\frac{1}{2} Y_5(t) \\
 \frac{1}{2}Y_3(t) &
   -\frac{1}{\sqrt{2}}Y_1(t)-\frac{1}{\sqrt{6}} Y_2(t)& -\frac{1}{2}
   Y_4(t) \\
 \frac{1}{2} Y_5(t)& -\frac{1}{2} Y_4(t) & \sqrt{\frac{2}{3}}
   Y_2(t)
\end{array} \right)~.
\end{eqnarray}
The $\so(1,2;\mathbb{R})$ Lie-algebra-valued connection  is then
obtained from the Lax operator with the standard $R$-matrix rule
\begin{eqnarray}\label{WEconn}
    W(t) & \equiv & L_{>}(t) - L_{<}(t) \nonumber\\
    & = & \left(
\begin{array}{lll}
 0 & -\frac{1}{2} Y_3(t) & -\frac{1}{2} Y_5(t) \\
 -\frac{1}{2} Y_3(t) & 0 & -\frac{1}{2} Y_4(t) \\
 -\frac{1}{2} Y_5(t) & \frac{1}{2} Y_4(t)& 0
\end{array}
\right)
\end{eqnarray}
and duely satisfies the condition
\begin{equation}\label{pseudorothogonalW}
    \eta \, W^T(t) + W(t)\, \eta \, = \, 0~.
\end{equation}
The Lax propagation equation is normalized as follows:
\begin{equation}\label{laxequa}
    \frac{d}{dt} \, L(t) \, + \, \left [ \, W(t) \, , \, L(t) \right ] \, =
    \,0~.
\end{equation}
Decomposed along the basis of coset generators $\mathrm{K}_i$ eq.(\ref{laxequa}) yields the following system of
five differential equations:
\begin{eqnarray}
  -\frac{Y_3(t)^2}{\sqrt{2}}-\frac{Y_4(t)^2}{2
   \sqrt{2}}-\frac{Y_5(t)^2}{2 \sqrt{2}}+\frac{d}{dt} \,Y_1(t) &=& 0~, \nonumber\\
  -\frac{1}{2} \sqrt{\frac{3}{2}} Y_4(t)^2+\frac{1}{2}
   \sqrt{\frac{3}{2}} Y_5(t)^2+\frac{d}{dt} \,Y_2(t) &=& 0~, \nonumber\\
-\sqrt{2} Y_1(t) Y_3(t)-Y_4(t) Y_5(t)+\frac{d}{dt} \,Y_3(t) &=& 0~, \nonumber\\
  \frac{Y_1(t)
   Y_4(t)}{\sqrt{2}}+\sqrt{\frac{3}{2}} Y_2(t)
   Y_4(t)-Y_3(t) Y_5(t)+\frac{d}{dt} \,Y_4(t) &=& 0~, \nonumber\\
  -\frac{Y_1(t)
   Y_5(t)}{\sqrt{2}}+\sqrt{\frac{3}{2}} Y_2(t)
   Y_5(t)+\frac{d}{dt} \,Y_5(t) &=& 0~. \label{diffequeLE}
\end{eqnarray}
This is the first order differential system on the $5$-dimensional Poissonian manifold
provided by the Borel subalgebra of $\mathrm{B}(\slal(3)) \subset \slal(3)$ which, as a consequence of the general discussion of section \ref{borellus}, is Liouville integrable. On this $5$-dimensional manifold there are four hamiltonian functions in involution. Of these latter one is
just zero, one is a Casimir, that labels the $4$-dimensional symplectic leaves into which the
$5$-dimensional manifold foliates. The remaining two functions are the $2$-hamiltonians which guarantee the Liouville integrability of each $4$-dimensional symplectic leaf. In the next section we consider the explicit form of such hamiltonians.
\subsection{The hamiltonian functions in involution}
As recalled above the differential system (\ref{diffequeLE}) is Liouville integrable since it admits
four hamiltonian functions in involution, that can be explicitly constructed according to formula (\ref{Hamiltonians}).
The first of these four hamiltonians  is identically zero since it corresponds to the trace of the Lax operator, namely to the extra generator of $\mathrm{gl(3,\mathbb{R})}$ which is deleted in order to step down to $\slal(3,\mathbb{R})$. The last, which, instead of being polynomial, is rational,  corresponds to the advertised Casimir labeling the leaves of the foliation.
\par
Explicitly from (\ref{tuttehamile}) we obtain the following result:
\begin{eqnarray}
\mathfrak{h}_1  \, \doteq  \, \mathfrak{h}_{01}  &=& 0~,\label{h1Ham} \\
 \mathfrak{h}_2 \, \doteq  \, \mathfrak{h}_{02}   &=& \frac{1}{2} Y_1(t)^2+\frac{1}{2}
   Y_2(t)^2-\frac{1}{4} Y_3(t)^2+\frac{1}{4}
   Y_4(t)^2-\frac{1}{4} Y_5(t)^2~,\label{h2Ham} \\
\mathfrak{h}_3 \, \doteq  \,  \mathfrak{h}_{03}  &=& \frac{Y_2(t)^3}{3 \sqrt{6}}-\frac{Y_1(t)^2
   Y_2(t)}{\sqrt{6}}+\frac{Y_3(t)^2 Y_2(t)}{2
   \sqrt{6}}+\frac{Y_4(t)^2 Y_2(t)}{4
   \sqrt{6}}-\frac{Y_5(t)^2 Y_2(t)}{4
   \sqrt{6}}\nonumber\\
   && -\frac{Y_1(t) Y_4(t)^2}{4
   \sqrt{2}}-\frac{Y_1(t) Y_5(t)^2}{4
   \sqrt{2}}+\frac{1}{4} Y_3(t) Y_4(t) Y_5(t)~,\label{h3Ham}\\
\mathfrak{h}_4 \, \doteq  \,  \mathfrak{h}_{11}  &=&
\frac{Y_1(t)}{\sqrt{2}}+\frac{Y_2(t)}{\sqrt{6}
   }-\frac{Y_3(t) Y_4(t)}{2 Y_5(t)}\label{h4Ham}
\end{eqnarray}
and by explicit calculation we can verify that the functions
$\mathfrak{h}_A$ ($A=1,\dots,4$) are constant along the Toda flow,
namely, upon use of eq.s(\ref{diffequeLE}) it is identically true
that
\begin{equation}\label{costantidelmoto}
    \partial_t \, \mathfrak{h}_A \, = \, 0~.
\end{equation}
If we calculate the secular equation for the Lax operator
(\ref{LElax}) we get
\begin{eqnarray}\label{secular}
    0 \, = \, \mbox{Det} \left ( L(t) \, - \, \lambda \, \mathbf{1}\right)\nonumber\\
     \, = \, -\lambda ^3+\mathfrak{h}_2 \lambda +\mathfrak{h}_3~.
\end{eqnarray}
Hence parameterizing the Toda flows by means of the values of the
hamiltonians, the eigenvalues of the Lax operator are given by the
three roots of the cubic equation  (\ref{secular}). Using
Cardano's formula these three roots are given by
\begin{eqnarray}
  \lambda_1 &=& -\frac{2 \sqrt[3]{3} \mathfrak{h}_2+\sqrt[3]{2}
   \left(\sqrt{\Delta }-9
   \mathfrak{h}_3\right)^{2/3}}{6^{2/3}
   \sqrt[3]{\sqrt{\Delta }-9 \mathfrak{h}_3}}~, \nonumber \\
  \lambda_2 &=& \frac{2 \left(3 \mathrm{i}+\sqrt{3}\right)
   \mathfrak{h}_2+\sqrt[3]{2} \sqrt[6]{3} \left(1-\mathrm{i}
   \sqrt{3}\right) \left(\sqrt{\Delta }-9
   \mathfrak{h}_3\right)^{2/3}}{2 2^{2/3} 3^{5/6}
   \sqrt[3]{\sqrt{\Delta }-9 \mathfrak{h}_3}}~, \nonumber\\
 \lambda_3 &=& \frac{2 \left(-3 \mathrm{i}+\sqrt{3}\right)
   \mathfrak{h}_2+\sqrt[3]{2} \sqrt[6]{3} \left(1+\mathrm{i}
   \sqrt{3}\right) \left(\sqrt{\Delta }-9
   \mathfrak{h}_3\right)^{2/3}}{2 2^{2/3} 3^{5/6}
   \sqrt[3]{\sqrt{\Delta }-9 \mathfrak{h}_3}}\label{autovalori}
\end{eqnarray}
where
\begin{equation}\label{discriminante}
    \Delta \, \equiv \, -12 \mathfrak{h}_2^3 \, + \, 81 \,
    \mathfrak{h}_3^2
\end{equation}
is the discriminant of the cubic equation.
\subsection{Normal form of the Lax operator}
The discussion of the normal forms  for the coset
$\frac{\mathrm{SL(3;\mathbb{R})}}{\mathrm{SO(1,2;\mathbb{R})}}$
splits in two cases since, from the definition of the number $p$
we have
\begin{equation}\label{pdefi}
    p \, \equiv \, \mbox{rank}\left (\frac{\mathrm{H}^\star}{\mathrm{H}_c} \right) \, = \,
    \mbox{rank}\left (\frac{\mathrm{SO(1,2;\mathbb{R})}}{\mathrm{SO(2;\mathbb{R})}} \right) \, = \,
    1~.
\end{equation}
Indeed the embedding (\ref{triaembed}) is trivial in this case
\begin{eqnarray}
\mathbb{U} \equiv \mathbb{ \slal(\mathrm{3},\mathbb{R})} & \hookrightarrow & \slal(\mathrm{3},\mathbb{R})~,\nonumber\\
\mathbb{U} \, \supset \, \mathbb{H}^\star \equiv
\so(1,2;\mathbb{R}) \,  & \hookrightarrow &
\so\mathrm{(1,3-1;\mathbb{R})} \, \subset \,
\slal(\mathrm{3},\mathbb{R})~. \label{triaembedtriv}
\end{eqnarray}
The two possible spectral type are characterized by $k=0$ or $k=1$.
\subsubsection{Spectral type $k=0$}
In this case we have from eq. (\ref{spectraltype}) $\mathcal{T} =
\, \widehat{T}_0 \, \,=\, \mathbf{1}_{3}$  and the normal form of
the Lax operator is
\begin{equation}\label{Lk0}
    \widehat{\Lambda}_{0} \, = \, \Lambda_{0} \, = \, \left(
                                                \begin{array}{ccc}
                                                  \lambda_1 & 0 & 0 \\
                                                  0 & \lambda_2 & 0 \\
                                                  0 & 0 & -\lambda_1-\lambda_2 \\
                                                \end{array}
                                              \right)
\end{equation}
where $\lambda_{1,2}$ are two real eigenvalues and
$-\lambda_1-\lambda_2$ is the third, also real. From the point of
view of the Lie algebra, the normal form (\ref{Lk0}) is just a
linear combination of the two Cartan generators
$\mathrm{K}_{1,2}$. Indeed we can write
\begin{equation}\label{K1K2Lam}
     \widehat{\Lambda} \, = \, \frac{1}{\sqrt{2}}\,\left(\lambda_1 -\lambda_2\right) \, \mathrm{K}_1 \, - \, \sqrt{\frac{3}{2}} \, \left (\lambda_1 + \lambda_2\right) \,
     \mathrm{K}_2~.
\end{equation}
A generic element of the $\mathrm{SO(1,2;\mathbb{R})}$ group can
be parameterized as a product of three elements of the three
one-parameter subgroups, namely we can set
\begin{eqnarray}\label{Ogeneric}
 &\mathrm{SO(1,2;\mathbb{R})} \, \ni \,  \mathcal{O}(v,\theta,w) = \exp \left [ v \, \mathrm{J_1}\right ]\,\cdot \,
 \exp \left [ \theta \, \mathrm{J_2}\right ] \,\cdot \, \exp \left [ w \, \mathrm{J_3}\right ]\, = \, &\nonumber\\
 & = {\scriptsize
 \left(
\begin{array}{lll}
 \cosh (v) \cosh (w)-\sin (\theta ) \sinh (v) \sinh
   (w) & -\cos (\theta ) \sinh (v) & \cosh (w) \sin
   (\theta ) \sinh (v)-\cosh (v) \sinh (w) \\
 \cosh (v) \sin (\theta ) \sinh (w)-\cosh (w) \sinh
   (v) & \cos (\theta ) \cosh (v) & \sinh (v) \sinh
   (w)-\cosh (v) \cosh (w) \sin (\theta ) \\
 -\cos (\theta ) \sinh (w) & \sin (\theta ) & \cos
   (\theta ) \cosh (w)
\end{array}
\right)}&\nonumber\\
\end{eqnarray}
where the parameters $v,w$ and $\theta$ can be regarded as the
three Euler angles (two hyperbolic and one elliptic) that
parameterize $\mathrm{SO(1,2;\mathbb{R})}$. Hence in the spectral
type $k=0$ the initial value of the Lax operator at \textit{time}
$t=0$ can be written as
\begin{equation}\label{L00}
    L_{0}(0) \, = \, \mathcal{O}_0 \cdot \widehat{\Lambda}_0 \cdot \mathcal{O}_0^{-1}
\end{equation}
where
\begin{equation}\label{Odefinition}
    \mathcal{O}_0 \, \equiv \, \mathcal{O}(v,\theta,w)~.
\end{equation}
In this way the matrix $L_{0}(0)$ depends on the five real parameters $\{\lambda_1,\lambda_2,v,\theta ,w\}$ which parameterize the initial conditions $Y_i(0)$ for the five real fields $Y_i(t)$. Indeed the values $Y_i(0)$
as functions of $\{\lambda_1,\lambda_2,v,\theta ,w\}$ can be extracted by projecting $L_{0}(0)$ along the orthonormal basis of coset generators $\mathrm{K}_i$.
\subsubsection{Spectral type $k=1$}
In this case we have from eq. (\ref{spectraltype})
\begin{equation}\label{Tk1}
    \mathcal{T} \, = \, \widehat{T}_1 \, \equiv \, \left(
\begin{array}{lll}
 \frac{1}{\sqrt{2}} & -\frac{i}{\sqrt{2}} & 0 \\
 -\frac{i}{\sqrt{2}} & \frac{1}{\sqrt{2}} & 0 \\
 0 & 0 & 1
\end{array}
\right)
\end{equation}
and the eigenvalues of the Lax operator are given by a pair of
complex conjugate eigenvalues $\lambda_1=x+{\rm i}y$, $\lambda_2 =
x-{\rm i}y$, while the third one is the negative of their sum,
$\lambda_3 = -2\,x$. Hence the normal form of the Lax operator is
as follows:
\begin{eqnarray}\label{Lh1}
    \widehat{\Lambda}_1 & = & \mathcal{T} \cdot \Lambda_1 \cdot \mathcal{T}^{-1} \, \equiv \, \left(
\begin{array}{lll}
 x & -y & 0 \\
 y & x & 0 \\
 0 & 0 & -2 x
\end{array}
\right)~, \nonumber\\
    \Lambda_1 & = & \left(
                      \begin{array}{ccc}
                        x+{\rm i}y & 0 & 0 \\
                        0 & x-{\rm i}y & 0 \\
                        0 & 0 & -\,2x \\
                      \end{array}
                    \right)~.
\end{eqnarray}
As element of the Lie algebra, rather then being a linear
combination of the two Cartan generators, in this case, the normal
form is a linear combination of one Cartan and one of the compact
coset generators. Indeed we have
\begin{equation}\label{frescobaldo}
    \widehat{\Lambda}_1 \, = \, -\sqrt{6} \, x \, \mathrm{K}_2 \, -\sqrt{2} \, y \,
    \mathrm{K}_3~.
\end{equation}
Hence in the spectral type $k=1$ the initial value of the Lax
operator at \textit{time} $t=0$ is written as
\begin{equation}\label{L10}
    L_{1}(0) \, = \, \mathcal{O}_0 \cdot \widehat{\Lambda}_1 \cdot \mathcal{O}_0^{-1}
\end{equation}
the rotation matrix $\mathcal{O}_0$ being defined in eq.(\ref{Odefinition}). As a result the initial values
$Y_i(0)$ of the $5$ fields are now parameterized by the five parameters $\{x,y,v,\theta,w\}$.
\par
In both spectral types we can calculate the values of the constant hamiltonians $\mathfrak{h}_A$ as functions of the five real parameter set, either $\{\lambda_1,\lambda_2,v,\theta,w\}$ or $\{x,y,v,\theta,w\}$. This is what we do in the next
section.
\subsection{Characterization of orbits through the hamiltonians}
It is now instructive to characterize the orbits and hence the
normal forms of the Lax operators through the values of the
conserved hamiltonians responsible for the integrability of the
system.
\par
As we have seen the two hamiltonians entering the secular equation
and hence the determination of the eigenvalues are
$\mathfrak{h}_{2,3}$. Furthermore, what distinguishes the two
spectral types is the sign of the discriminant $\Delta$ defined in
equation (\ref{discriminante}). When $\Delta < 0$ we have three
real eigenvalues, while when $\Delta > 0$ we have a pair of
complex conjugate eigenvalues and a third real one. The regions in
the $\mathfrak{h}_2,\mathfrak{h}_3$ plane corresponding to the two
spectral types are visualized in Fig.\ref{phasediagr} \iffigs
\begin{figure}
\begin{center}
\includegraphics[height=9cm]{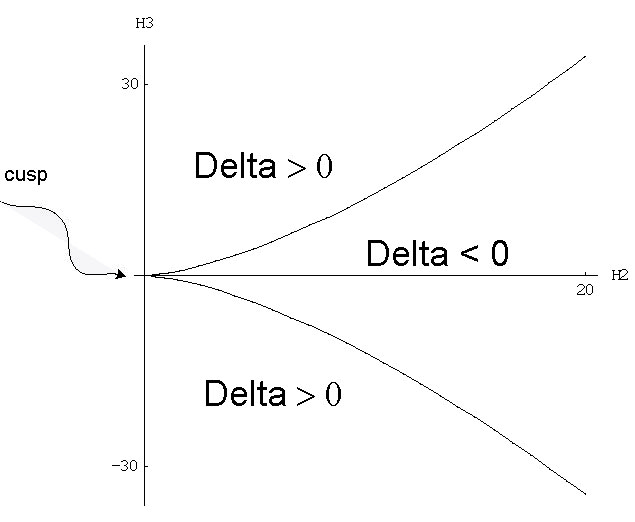}
\end{center}
\else\fi \caption{\label{phasediagr} Phase-diagram in the
$\mathfrak{h}_2,\mathfrak{h}_3$ plane. In the locus $\Delta > 0$
we have complex conjugate eigenvalues and spectral type $k=1$. In
the locus $\Delta < 0$ there are three real eigenvalues and
spectral type $k=0$. The line $\Delta=0$ which separates the two
regions is a singular locus where two eigenvalues coincide and we
have an enhancement of symmetry. On this locus the normal form
admits a one-parameter stability subgroup. The cuspidal point
$\mathfrak{h}_2 =\mathfrak{h}_3 \, = \,0$ corresponds to nilpotent
Lax operators as we discuss in the paper \cite{noisecondo}
submitted to the hep-th arXiv after the first appearance of the
present paper.}
\end{figure}
It is now instructive to evaluate the explicit form of the
hamiltonians and of the discriminant in the two spectral types.
\subsubsection{The hamiltonians in spectral type $k=0$}
>From the definition given in eq.s(\ref{h1Ham}-\ref{h4Ham}), by
using the initial Lax operator (\ref{L00}) to calculate the fields
$Y_i(0)$ and hence the hamiltonians, we find
\begin{eqnarray}
  \mathfrak{h}_1 &=& 0~, \label{h10}\\
  \mathfrak{h}_2 &=& \lambda _1^2+\lambda _2 \lambda _1+\lambda _2^2~, \label{h20}\\
  \mathfrak{h}_3 &=& -\lambda _1 \lambda _2 \left(\lambda _1+\lambda
   _2\right)~, \label{h30}\\
  \mathfrak{h}_4 &=& -\left[2 \left(\sin \theta  \sinh v \lambda
   _1^2+2 \cosh w (\cosh w \sin \theta  \sinh
   v-\cosh v \sinh w) \lambda _2 \lambda
   _1  \right. \right. \nonumber\\
   &&\left.\left. +\sinh w (\sin \theta  \sinh v \sinh
   w-\cosh v \cosh w) \lambda _2^2\right)\right]  \nonumber\\
   && \times \, \left[2
   \left(\sin \theta  \sinh v \cosh ^2w-2
   \cosh v \sinh w \cosh w+\sin\theta
   \sinh v \sinh ^2w\right) \lambda _1 \right. \nonumber\\
   &&\left.+((\cosh
   2 w+3) \sin \theta  \sinh v-2 \cosh v
   \cosh w \sinh w) \lambda _2\right]^{-1}~,\label{h40}\\
  \Delta &=& -3 \left(\lambda _1-\lambda _2\right)^2 \left(2
   \lambda _1+\lambda _2\right)^2 \left(\lambda _1+2
   \lambda _2\right)^2~.
\end{eqnarray}
As it is evident from the above explicit expressions, in the
spectral type $k=0$ the discriminant is strictly negative and it
reaches the value zero only in the case of degenerate eigenvalues,
namely when any two of the three eigenvalues are equal.
\subsubsection{The hamiltonians in the spectral type $k=1$}
>From the definition given in eq.s(\ref{h1Ham}-\ref{h4Ham}), by
using the initial Lax operator (\ref{L10}) to calculate the fields
$Y_i(0)$ and hence the hamiltonians, we find
\begin{eqnarray}
  \mathfrak{h}_1 &=& 0~, \label{h11}\\
  \mathfrak{h}_2 &=& 3 x^2-y^2~,\label{h12}\\
  \mathfrak{h}_3 &=& -2 x \left(x^2+y^2\right)~, \label{h13}\\
  \mathfrak{h}_4 &=& \left[\cosh v \left(\cos \theta  \sinh 2 w
   y^2+2 x \cosh w (2 y \sin \theta +3 x \cos
   \theta  \sinh w)\right)-\right.\nonumber\\
   &&\left. \sinh v
   \left(\left(3 x^2+y^2\right) \cos \theta  \sin
   \theta  \cosh ^2 w+\left(3 x^2+y^2\right) \cos
   \theta  \sin \theta  \sinh ^2 w \right. \right.\nonumber\\
   &&\left.\left. +\left(3
   x^2+y^2\right) \cos \theta  \sin \theta -4 x
   y \cos 2 \theta  \sinh w\right)\right] \, \times \nonumber\\
   && \times \, \left[2
   \left(\frac{3}{2} x \cos \theta  \sin \theta
   \sinh v \cosh ^2 w +y \cosh v \sin \theta
   \cosh w \right.\right.\nonumber\\
   &&\left.\left.-3 x \cos \theta  \cosh v \sinh w
   \cosh w+\frac{3}{2} x \cos \theta  \sin
   \theta  \sinh v \sinh ^2 w\right.\right.\nonumber\\
   &&\left.\left.+\frac{3}{2} x
   \cos \theta  \sin \theta  \sinh v+y \cos (2
   \theta ) \sinh v \sinh w\right)\right]^{-1}~, \label{h14}\\
  \Delta &=& 12 y^2 \left(9 x^2+y^2\right)^2~. \label{Deltone1}
\end{eqnarray}
Once again also in this branch of the solution space the sign of the discriminant is definite. $\Delta$ is positive definite and it vanishes only for $y=0$. This condition however corresponds to real degenerate eigenvalues and matches the same condition obtained from the other branch with spectral type $k=0$.
\section{Examples of explicit solutions}
In this section we illustrate the integration algorithm by considering some example of solutions corresponding to the three spectral types: $k=0$, $k=1$ and degenerate.
\subsection{An example of solution of spectral type $k=0$}
A very simple solution of this spectral type can be obtained fixing
the following initial data:
\begin{eqnarray}
  \{\lambda_1,\lambda_2,\lambda_3 \}&=& \{ \ft 12 \, , \, \ft 32 \, , \,  -2\} ~,\\
  \mathcal{O}_0 &=& \exp[2\, \mathrm{J}_1]  \, = \, \left(
\begin{array}{lll}
 \frac{1+e^4}{2 e^2} & -\frac{-1+e^4}{2 e^2} & 0 \\
 -\frac{-1+e^4}{2 e^2} & \frac{1+e^4}{2 e^2} & 0 \\
 0 & 0 & 1
\end{array}
\right)\quad \quad (v=2,\theta=0,w=0).\nonumber\\
\end{eqnarray}
With these data the initial form of the Lax operator is the following:
\begin{equation}\label{iniL001}
    L_0(0) \, = \, O_0 \cdot \left(
                               \begin{array}{ccc}
                                 \ft 12 & 0 & 0 \\
                                 0 & \ft 32 & 0 \\
                                 0 & 0 & -2 \\
                               \end{array}
                             \right) \cdot \, O_0^{-1} \, = \, \left(
\begin{array}{lll}
 1-\frac{1}{4 e^4}-\frac{e^4}{4} & -\frac{-1+e^8}{4
   e^4} & 0 \\
 \frac{-1+e^8}{4 e^4} & \frac{1}{4}
   \left(4+\frac{1}{e^4}+e^4\right) & 0 \\
 0 & 0 & -2
\end{array}
\right)~.
\end{equation}
As we see $L_0(0)$ has a  block diagonal structure $2+1$. Such a
block structure is preserved throughout the all flow from
$t=-\infty$ to $t=+\infty$ as we can deduce from the explicit
result of the integration
\begin{eqnarray}
  Y_1(t) &=& -\frac{1-2 e^4+e^8+e^{2 t}+2 e^{2 t+4}+e^{2
   t+8}}{\sqrt{2} \left(-1+2 e^4-e^8+e^{2 t}+2 e^{2
   t+4}+e^{2 t+8}\right)}~, \\
  Y_2(t) &=& -\sqrt{6}~, \\
  Y_3(t) &=& -\frac{2 e^t \left(-1+e^8\right)}{-1+2 e^4-e^8+e^{2
   t}+2 e^{2 t+4}+e^{2 t+8}}~,\\
  Y_4(t) &=& 0~, \\
  Y_5(t) &=& 0~.
\end{eqnarray}
The vanishing of both $Y_4(t)$ and $Y_5(t)$ is what guarantees the block diagonal structure of the Lax operator. The same fact however implies that the $4$-th hamiltonian, the rational one is indeterminate in this case, being the ratio of two zeros. The other two (polynomial) hamiltonians have instead the following explicit values:
\begin{equation}\label{fuffo}
    \mathfrak{h}_2 \, = \, \frac{13}{4} \quad ; \quad \mathfrak{h}_3 \, = \,
    -\frac{3}{2}~.
\end{equation}
The plot of the two non-trivial functions $Y_{1,3}(t)$ is
exhibited in Fig.\ref{Y13}. As we see there is a singularity in
both fields at a finite time $t = t_0 \, \simeq -0.03663$. \iffigs
\begin{figure}
\begin{center}
\includegraphics[height=12cm]{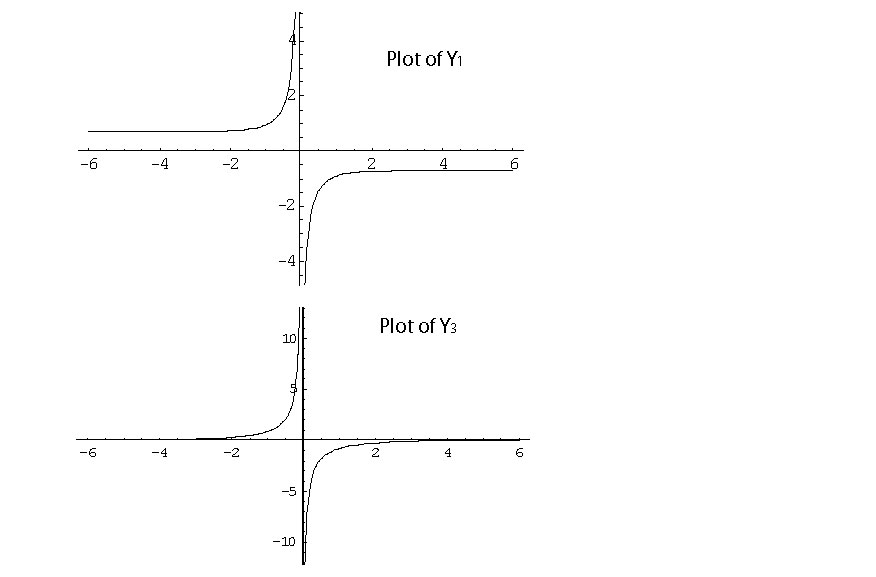}
\end{center}
\else\fi
\caption{\label{Y13} Plot of the $Y_{1,3}(t)$ fields in the solution of spectral type $k=0$ characterized by the following initial data: $\lambda_1=\ft 12$, $\lambda_2 \, = \, \ft 32$, $v=2$, $\theta = w = 0$. In both fields
there is a singularity at $t=t_s \simeq -0.03663$. The other three fields are constant or even zero.}
\end{figure}
This singularity separates the range of the variable $t$ in two
parts. We can consider the solution only on one side of the
singularity. Let us consider for instance the Cartan field
$Y_1(t)$. As we know this function is actually the derivative of
the corresponding Cartan field $h_1(t)$ \cite{Fre':2007hd} and in
order to reconstruct the physical interpretation of our solution
we are supposed to perform a second integration
\begin{equation}\label{cartanus}
    h_1(t) \, = \, \int \, Y_{1}(t)\, dt \, = \, \frac{t-\log \left|-1+2 e^4-e^8+e^{2 t}+2 e^{2
   t+4}+e^{2 t+8}\right|}{\sqrt{2}}~.
\end{equation}
\par
The singularity at a finite time is a qualitative difference
between the type of solutions encountered in the pseudo-Riemannian
case and those encountered in the case of cosmic billiards
(Riemannian coset manifolds). A similarity, instead, which exists
between the spectral type $k=0$ of pseudo-Riemannian system and
the billiard case is the asymptotic behavior at $t=\pm \infty$.
For this spectral type (but not for the other), just as in the
billiard case,  the Lax operator tends asymptotically to a
diagonal form which differs from $\Lambda$ only by a permutation
of the eigenvalues. We can verify this statement in the present
example. We find
\begin{eqnarray}
  L_0(-\infty) &=& \left(
\begin{array}{lll}
 \frac{3}{2} & 0 & 0 \\
 0 & \frac{1}{2} & 0 \\
 0 & 0 & -2
\end{array}
\right) \, \ne \, \Lambda ~,\nonumber \\
  L_0(\infty) &=& \left(
\begin{array}{lll}
 \frac{1}{2} & 0 & 0 \\
 0 & \frac{3}{2} & 0 \\
 0 & 0 & -2
\end{array}
\right) \, = \, \Lambda~.
\end{eqnarray}
\subsection{Another solution of the  spectral type $k=0$ with finite hamiltonians}
To appreciate the differences we apply the integration algorithm
to the case where the choice of the eigenvalues and of the
spectral type remains the same as in the previous example but we
modify the initial rotation element $\mathcal{O}_0$ by switching
on also a compact rotation angle $\theta = \frac{\pi}{4}$. So we
set
\begin{equation}\label{newO}
    \mathcal{O}_0 \, = \, \exp\left [2 \,\mathrm{J}_1\right ]\,\cdot\, \exp\left [ \frac{\pi}{4} \mathrm{J}_2\right] \, = \, \left(
\begin{array}{lll}
 \frac{1+e^4}{2 e^2} & -\frac{-1+e^4}{2 \sqrt{2} e^2}
   & \frac{-1+e^4}{2 \sqrt{2} e^2} \\
 -\frac{-1+e^4}{2 e^2} & \frac{1+e^4}{2 \sqrt{2} e^2}
   & -\frac{1+e^4}{2 \sqrt{2} e^2} \\
 0 & \frac{1}{\sqrt{2}} & \frac{1}{\sqrt{2}}
\end{array}
\right)
\end{equation}
and for the initial Lax operator we get
\begin{equation}\label{L0primo}
    L_0(0) \, = \, \mathcal{O}_0 \, \Lambda \, \mathcal{O}_0^{-1} \, = \, \left(
\begin{array}{lll}
 \frac{3+2 e^4+3 e^8}{16 e^4} & \frac{3
   \left(-1+e^8\right)}{16 e^4} & -\frac{7
   \left(-1+e^4\right)}{8 e^2} \\
 -\frac{3 \left(-1+e^8\right)}{16 e^4} & -\frac{3-2
   e^4+3 e^8}{16 e^4} & \frac{7 \left(1+e^4\right)}{8
   e^2} \\
 \frac{7 \left(-1+e^4\right)}{8 e^2} & \frac{7
   \left(1+e^4\right)}{8 e^2} & -\frac{1}{4}
\end{array}
\right)~.
\end{equation}
Calculating the hamiltonians from the above form of the initial
Lax operator we obtain
\begin{equation}\label{finitehamilte}
 \{ \mathfrak{h}_1,\mathfrak{h}_2,\mathfrak{h}_3,\mathfrak{h}_4\} \, = \, \left\{0,\frac{13}{4},-\frac{3}{2},-\frac{1}{2}\right
   \}~.
\end{equation}
As we see $\mathfrak{h}_2$ and $\mathfrak{h}_3$, which depend only on the eigenvalues are the same as before. On the other hand, $\mathfrak{h}_4$  is no longer undefined as in the previous case and obtains the finite rational value $-\ft 12$. This is so because the new initial value of the Lax operator as no degenerate minors has in the previous case.
\par
The new explicit solution is given by the following functions:
{\scriptsize
\begin{eqnarray}
  Y_1(t) &=& -\frac{\left(-2+5 e^{7 t}\right) \left(1-2 e^4+e^8+2
   e^{2 t}+e^{7 t}+4 e^{2 t+4}+2 e^{2 t+8}-2 e^{7
   t+4}+e^{7 t+8}\right)}{2 \sqrt{2} \left(1+e^{7
   t}\right) \left(1-2 e^4+e^8-2 e^{2 t}+e^{7 t}-4
   e^{2 t+4}-2 e^{2 t+8}-2 e^{7 t+4}+e^{7
   t+8}\right)}~, \\
  Y_2(t) &=& \frac{\sqrt{\frac{3}{2}} \left(-4+3 e^{7 t}\right)}{2
   \left(1+e^{7 t}\right)}~, \\
  Y_3(t) &=& -\frac{\sqrt{2} e^t \left(-1+e^8\right) \left(-2+5
   e^{7 t}\right)}{\sqrt{1+e^{7 t}} \left(1-2
   e^4+e^8-2 e^{2 t}+e^{7 t}-4 e^{2 t+4}-2 e^{2
   t+8}-2 e^{7 t+4}+e^{7 t+8}\right)}~, \\
  Y_4(t) &=& -\frac{7 \sqrt{2} e^{-t-2}
   \left(1+e^4\right)}{\sqrt{-e^{-11 t-4}
   \left(1+e^{7 t}\right)^2 \left(1-2 e^4+e^8-2 e^{2
   t}+e^{7 t}-4 e^{2 t+4}-2 e^{2 t+8}-2 e^{7
   t+4}+e^{7 t+8}\right)}}~, \\
  Y_5(t)&=& -\frac{7 \left(-1+e^4\right)}{e^2 \sqrt{-e^{-7 t-4}
   \left(1+e^{7 t}\right) \left(1-2 e^4+e^8-2 e^{2
   t}+e^{7 t}-4 e^{2 t+4}-2 e^{2 t+8}-2 e^{7
   t+4}+e^{7 t+8}\right)}}~.
\end{eqnarray}
} The behavior of the solution is qualitatively similar to that
discussed in the previous case. This is evident from the plots of
the Cartan fields exhibited in Fig.\ref{Y12sol2}. \iffigs
\begin{figure}
\begin{center}
\includegraphics[height=12cm]{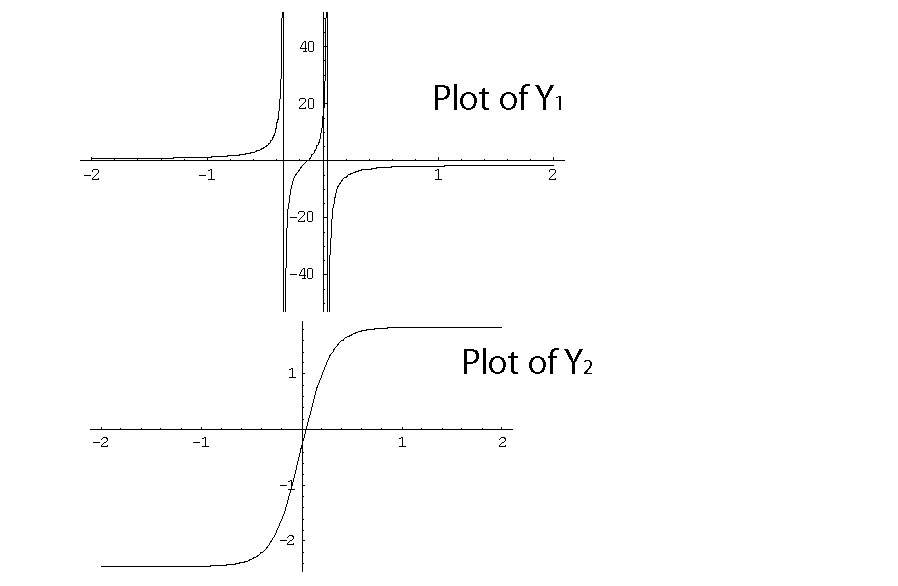}
\end{center}
\else\fi
\caption{\label{Y12sol2} Plot of the Cartan fields $Y_{1,2}(t)$  in the solution of spectral type $k=0$ characterized by the following initial data: $\lambda_1=\ft 12$, $\lambda_2 \, = \, \ft 32$, $v=2$, $\theta = \pi/4$, $ w = 0$. In $Y_1(t)$ there are two singularities at finite times $t=t_{s_1} \simeq -0.338507$
and $t_{s_2} \simeq 0.0417536$.}
\end{figure}
As one realizes there are just two singularities at finite time
$t=t_{s_1} \simeq -0.338507$ and $t_{s_2} \simeq 0.0417536$, that
affect one of the two Cartans but not the other. The same
singularities appear in the field $Y_{3}(t)$. The other two fields
$Y_{4,5}(t)$ are real only in the interval between the two
singularities as it evident from the plot of, for instance,
$Y_{4}(t)$, exhibited in Fig.\ref{Y4sol2}. \iffigs
\begin{figure}
\begin{center}
\includegraphics[height=6cm]{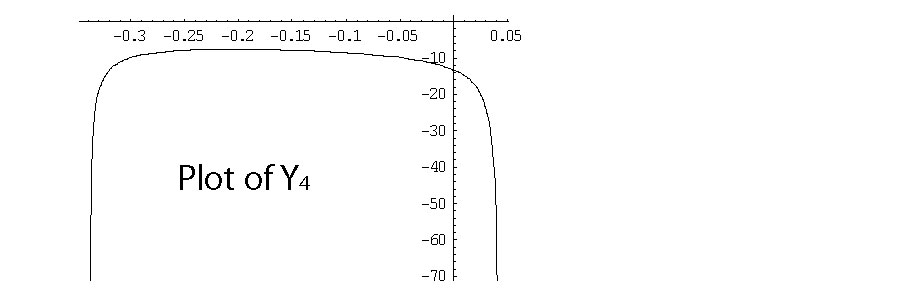}
\end{center}
\else\fi
\caption{\label{Y4sol2} Plot of the non-compact nilpotent fields $Y_{4}(t)$  in the solution of spectral type $k=0$ characterized by the following initial data: $\lambda_1=\ft 12$, $\lambda_2 \, = \, \ft 32$, $v=2$, $\theta = \pi/4$, $ w = 0$. The function $Y_4(t)$ is real only in the interval comprised
between the two singularities $t=t_{s_1} \simeq -0.338507$
and $t_{s_2} \simeq 0.0417536$.}
\end{figure}
This means that the overall solution is properly defined only in the interval between the two singularities.
Notwithstanding this fact if we calculate the asymptotic limit of the Lax operator at $\pm\infty$ we obtain
finite diagonal real forms. Indeed we find
\begin{eqnarray}
  L_0(-\infty) &=& \left(
\begin{array}{lll}
 \frac{3}{2} & 0 & 0 \\
 0 & \frac{1}{2} & 0 \\
 0 & 0 & -2
\end{array}
\right) \, \ne \, \Lambda~, \nonumber \\
  L_0(\infty) &=& \left(
\begin{array}{lll}
 -2 & 0 & 0 \\
 0 & \frac{1}{2} & 0 \\
 0 & 0 & \frac{3}{2}
\end{array}
\right) \, \ne \, \Lambda~.
\end{eqnarray}
In the present example the difference between the order of eigenvalues at $+\infty$ and at $-\infty$ is provided by the permutation of highest order, just as it happens in the billiard Riemannian case for flows not touching singular surfaces. The previous case did not have this property because it developed on a singular surface and the indeterminacy of the fourth hamiltonian was a sign of that.
\par
Although similar to the billiard case the asymptotic limits loose
their meaningfulness in the pseudo--Riemannian case since they are
separated from the physical flow region by regions where the Lax
operator becomes complex. The real asymptotic diagonal limits are
approached through imaginary values. The physical flow region is
typically bounded by singularities.
\subsection{An example of spectral type $k=1$}
As an example of the other spectral type we choose the solution
generated by the following very simple initial data:
\begin{eqnarray}
  k &=& 1 \quad x=1 \quad ; \quad y=1~, \\
  v &=& 0~, \\
  \theta &=& \frac{\pi}{3}~, \\
  w &=& 0~.
\end{eqnarray}
The corresponding rotation matrix is
\begin{equation}\label{rotamatrona}
    \mathcal{O}_0 \, = \, \left(
\begin{array}{lll}
 1 & 0 & 0 \\
 0 & \frac{1}{2} & -\frac{\sqrt{3}}{2} \\
 0 & \frac{\sqrt{3}}{2} & \frac{1}{2}
\end{array}
\right)
\end{equation}
and the resulting initial Lax operator is
\begin{equation}\label{frescusLax}
    L_1(0) \, = \, \left(
\begin{array}{lll}
 1 & -\frac{1}{2} & -\frac{\sqrt{3}}{2} \\
 \frac{1}{2} & -\frac{5}{4} & \frac{3 \sqrt{3}}{4} \\
 \frac{\sqrt{3}}{2} & \frac{3 \sqrt{3}}{4} &
   \frac{1}{4}
\end{array}
\right)~.
\end{equation}
The resulting vector of hamiltonians is
\begin{equation}\label{hamiltonAncor}
  \{ \mathfrak{h}_1,\mathfrak{h}_2,\mathfrak{h}_3,\mathfrak{h}_4\} \,   = \, \{0, 2, -4, 2\}
\end{equation}
and the explicit form of the solution is given by
\begin{eqnarray}
  Y_1(t) &=& \frac{\sec (2 t) \left(9 e^{6 t} \cos (4 t)+4 \sin (2
   t)+3 e^{6 t} (\sin (4 t)+3)\right)}{\sqrt{2}
   \left(6 e^{6 t} \cos (2 t)+2\right)}~,\\
  Y_2(t) &=& \frac{\sqrt{\frac{3}{2}} \left(3 e^{6 t} \cos (2 t)-3
   e^{6 t} \sin (2 t)-2\right)}{3 e^{6 t} \cos (2
   t)+1}~, \\
  Y_3(t) &=& -\frac{2 \sqrt{2} e^{-3 t}}{\cos ^{\frac{1}{2}}(2 t)
   \sqrt{2 e^{-6 t} \cos (2 t)+3 \cos (4 t)+3}}~, \\
  Y_4(t) &=& -\frac{2 \sqrt{6} e^{-2 t} (3 \cos (2 t)-\sin (2
   t))}{\sqrt{3 e^{2 t} \cos (2 t)+e^{-4 t}} \sqrt{2
   e^{-6 t} \cos (2 t)+3 \cos (4 t)+3}}~, \\
  Y_5(t) &=& -\frac{2 \sqrt{3} e^t}{\cos ^{\frac{1}{2}}(2 t)
   \sqrt{3 e^{2 t} \cos (2 t)+e^{-4 t}}}~.
\end{eqnarray}
The structure of this solution can be considered analyzing the
plots of the various fields. The Cartan fields exhibit a quasi
periodic behavior (with singularities) displayed in
fig.\ref{figurona4}. \iffigs
\begin{figure}
\begin{center}
\includegraphics[height=11cm]{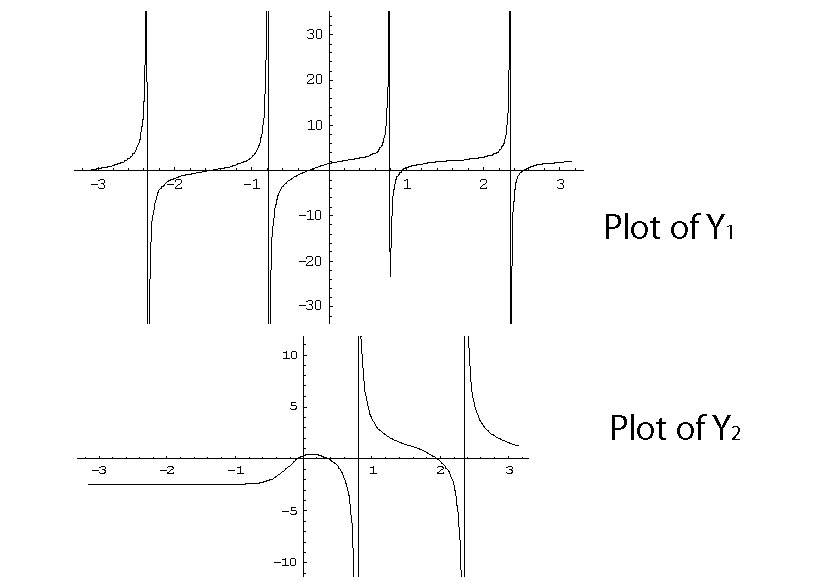}
\end{center}
\else\fi
\caption{\label{figurona4} Plot of the two Cartan fields $Y_{1,2}(t)$ in the solution of
spectral type $k=1$, with parameters $x=1,y=1,v=0,\theta=\frac{\pi}{3},w=0$.}
\end{figure}
The two fields $Y_{3,5}(t)$ have instead a periodic real behavior for $t>t_m$ and for $t<t_p$ respectively, where
\begin{equation}\label{tpnuovo}
    t_p \, \simeq \, 0.785398 \quad ; \quad t_m \, \simeq \, - 0.785398 \, = \, - t_p
\end{equation}
are finite  times. Respectively below and above these singularity barriers
 the fields $Y_{3,5}(t)$ become imaginary. This is evident from the plots displayed
 in Fig. \ref{figurona5}
\iffigs
\begin{figure}
\begin{center}
\includegraphics[height=11cm]{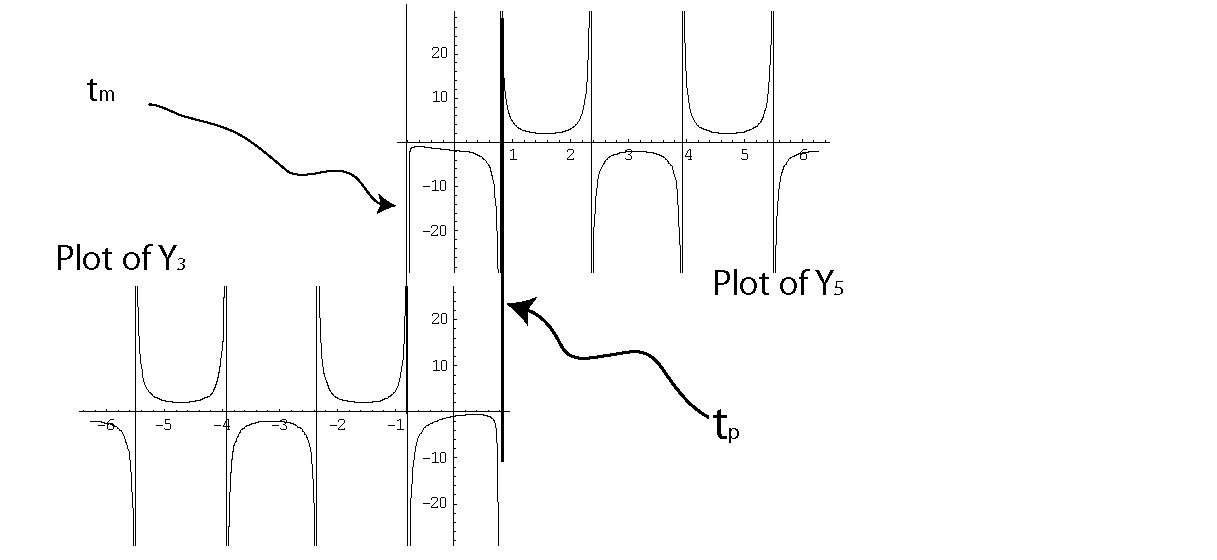}
\end{center}
\else\fi
\caption{\label{figurona5} Plot of the non  Cartan fields $Y_{3,5}(t)$ in the solution of
spectral type $k=1$, with parameters $x=1,y=1,v=0,\theta=\frac{\pi}{3},w=0$.}
\end{figure}
This behavior restricts the physical range of the solution to the
interval $[-t_p\, , \, t_p]$. This is further confirmed by the
plot of the field $Y_{4}(t)$ which is real only in the same
interval. This is seen in Fig.\ref{figurona6} \iffigs
\begin{figure}
\begin{center}
\includegraphics[height=9cm]{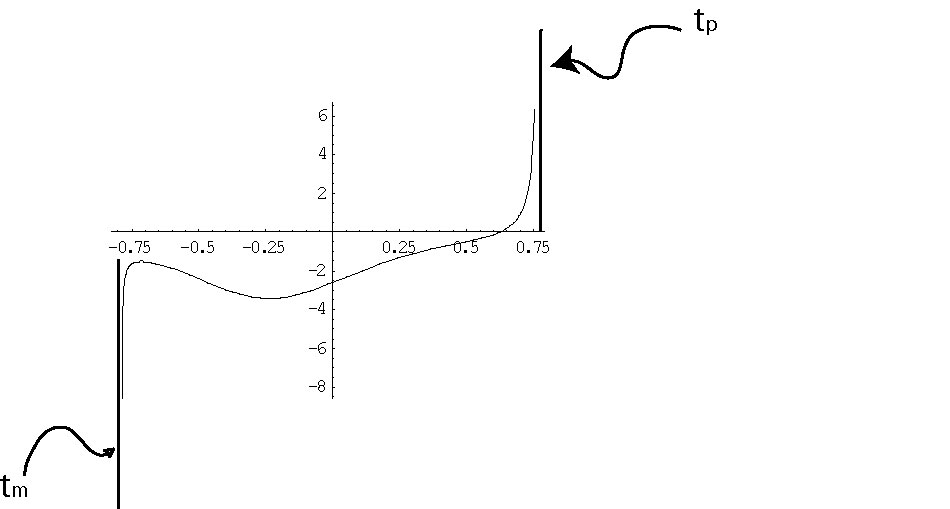}
\end{center}
\else\fi \caption{\label{figurona6} Plot of the non  Cartan field
$Y_{4}(t)$ in the solution of spectral type $k=1$, with parameters
$x=1,y=1,v=0,\theta=\frac{\pi}{3},w=0$. Outside of the plotted
range the field is imaginary.}
\end{figure}
\par
In this way we come to the conclusion that, notwithstanding the appearance of periodic functions and the periodic behavior of some of the fields of the system, also in the case of the spectral type $k=1$, the generic form of the real solution appears to be the evolution on a finite range of the \textit{time}-line, bounded at the extrema by singularities. A similar generic behavior is suitable for the description of an evolution from spatial infinity to a horizon as it happens in black hole physics.
\par
A detailed study of the solution space, a classification of the asymptotic limits and the analysis of critical surfaces in the moduli space is postponed to future publications, where the physical interpretation of the Lax equation solutions in connection with $p$-brane physics will be addressed.
\section{Conclusions}
In this paper we have presented a new view-point on the integrability of supergravity cosmic billiards and black holes
that is based on the Poissonian structure of the underlying solvable Lie algebra $\mathcal{S}$.
\par
The main  results of our paper are two:
\begin{itemize}
  \item The explicit construction of the integration algorithm extended also to the case of Lorentzian cosets
  $\mathrm{U/H}^\star$.
  \item The explicit construction of the hamiltonian functions in involution $\mathfrak{h}_\alpha$ responsible for Liouville
  integrability.
\end{itemize}
We believe that a systematic use of our techniques for the construction of black-hole and billiard solutions will provide new results and new insight. In particular the relation of the Hamiltonians and the Casimirs with the physical invariants of the solution, like the entropy or the total mass will prove very helpful and inspiring. We leave this to future coming publications.
\par
A point which we have not yet addressed but which is of the highest relevance concerns the issue of global topology of the solution space. The solvable parametrization covers only open branches of this space and the question of how to glue together different branches is very important.
\par
Also this issue is left over for future publications.
\appendix
\paragraph{Aknowledgments} We would like to express our gratitude to our frequent collaborator and excellent friend Mario Trigiante for the exchange of useful information we had with him at the end of this project. He attracted our attention
to the classification of normal forms already performed in \cite{Bergshoeff:2008be}, which coincides with the classification of spectral types we came up with in our adaptation of the Kodama integration algorithm. Furthermore we
have been informed that explicit solutions for the $\mathrm{SL(3,R)/SO(1,2)}$ case, although not yet published, were
already constructed more than a month ago by him and his collaborators, implementing their own adaptation of the Kodama algorithm. Although we have not yet seen these solutions we are absolutely confident that they will be coherent with our own results. It is our pleasure to aknowledge this fact publicly.
\vskip 0.3cm
\paragraph{Note added in the revised version} As stated in the previous aknowledgements, the authors of \cite{marioetal} had indeed independently adapted Kodama integration algorithm to the case of
 $\mathrm{G/H}^\star$ Lax equations and  had  derived some explicit solutions for the cases of $\mathrm{SL(2,\mathbb{R})/SO(1,1)}$ and $\mathrm{SL(3,\mathbb{R})/SO(1,2)}$. The construction of the integration algorithm for the case of nilpotent initial Lax operators, which is relevant for  extremal Black Holes, was
 performed in full generality in our paper \cite{noisecondo}. In a revised version of their paper, which appeared the
 same day as our \cite{noisecondo}, the authors of \cite{marioetal} presented some particular solutions corresponding to specific nilpotent Lax operators pertaining to $\mathrm{SL(3,\mathbb{R})/SO(1,2)}$.

\end{document}